\documentclass[12pt]{article}
\usepackage{cite}
\usepackage{axodraw}
\usepackage{lscape}
\begin{document}
\title{\bf Renormalization of the energy-momentum tensor in
           noncommutative complex scalar field theory}

\author{\sc S. Bellucci${}^{a}$,
I.L. Buchbinder${}^{b}$,
V.A. Krykhtin${}^c$\footnote{e-mail:
\tt bellucci@lnf.infn.it, joseph@tspu.edu.ru, krykhtin@mph.phtd.tpu.edu.ru}
\\[0.5cm]
\it ${}^a$INFN, Laboratori Nazionali di Frascati,\\
\it P.O. Box 13, I-00044 Frascati, Italy\\[0.3cm]
\it ${}^b$Department of Theoretical Physics,\\
\it Tomsk State Pedagogical University,\\
\it Tomsk 634041, Russia\\[0.3cm]
\it ${}^c$Laboratory of Mathematical Physics and\\
\it Department of Theoretical and Experimental Physics, \\
\it Tomsk Polytechnic University,\\
\it Tomsk 634050, Russia }

\date{}


\maketitle
\thispagestyle{empty}

\vspace{0.5cm}

\begin{abstract}
We study the renormalization of dimension four
composite operators and the energy-momentum tensor
in noncommutative complex scalar field theory.
The proper operator basis is defined and
it is proved that the bare
composite operators are expressed via renormalized ones with the
help of an appropriate mixing matrix which is calculated in the one-loop
approximation. The number and form of the operators in the basis and the
structure of the mixing matrix essentially differ from those in
 the corresponding commutative theory and in noncommutative real
scalar field theory. We show that the energy-momentum
tensor in the noncommutative complex scalar field theory is defined up
to six arbitrary constants.  The canonically defined energy-momentum
tensor is not finite and must be replaced by the "improved" one, in
order to provide finiteness.  Suitable "improving" terms are found.
Renormalization of
dimension four composite operators at zero momentum transfer is also
studied.  It is shown that the mixing matrices are different for the
cases of arbitrary and zero momentum transfer. The energy-momentum vector,
unlike the energy-momentum tensor,
is defined unambigously and does not require "improving", in order to be
conserved and finite, at least in the one-loop approximation.
\end{abstract}


\section{Introduction}

The study of noncommutative field theories has
attracted much attention lately, due to their profound links
with the string
theory \cite{9908142} and remarkable properties in classical
and quantum domains (see e.g. the reviews
\cite{0106048,0109162,0111208}).

There exist two general aspects of renormalization procedure in any
field theory. Firstly, the renormalization of Green functions or
effective action and secondly, the renormalization of composite operators
(see e.g.  \cite{Collins} for a discussion of this problem in
the commutative theories).  The problem of renormalization of Green's
functions (i.e.  fields, masses, and coupling constants) was studied
for many noncommutative field theories to different approximation
orders (see e.g.
\cite{9911098,9912072,9912075,0008057,0001215,0207086} and the reviews
\cite{0106048,0109162,0111208}).

The present paper is devoted to the problem of renormalization
of composite operators and
the energy-mo\-men\-tum tensor in noncommutative complex
scalar field theory.
The analogous problem in noncommutative real field
theory was considered in \cite{0303186}. As we will see, the
renormalization of composite operators in noncommutative complex
scalar field theory essentially differs from that in noncommutative real
field theory.\footnote{Problem of constructing the classical
energy-momentum tensor in noncommutative field theories is
discussed in
\cite{0012112,0103124,0104244,0210288,0212122,0310155}.}

A noncommutative field theory is usually constructed from the
corresponding commutative theory, by replacing the pointwise product
of the fields with the star one
\begin{eqnarray}
\label{theta}
f\cdot g\to
(f\star g)(x)=
  \left.\exp(\frac{i}{2}\theta^{\mu\nu}\partial_\mu^u\partial_\nu^v)
                f(x+u)g(x+v)\right|_{u=v=0} \neq (g\star f)(x),
\end{eqnarray}
where the constants $\theta^{\mu\nu}$ are noncommutativity
parameters with dimension of a length squared.
The star product (\ref{theta}) is noncommutative, so, in
contrast to the commutative field theories, there is a problem
of ordering of the fields in the Lagrangian of noncommutative
theory.
In the noncommutative real scalar field theory there was only
one kind of field and this problem was absent.
Therefore in that case both commutative and noncommutative
theories had one coupling constant.
In the case of noncommutative complex scalar field theory we
have two kinds of fields and the problem of fields ordering
arises.
Therefore we have to take into account all possible ways of field ordering.
The action of the theory is the Lagrangian integrated over the whole space-time.
However, when we integrate the star product (\ref{theta}) over the
whole space-time, we can prove, integrating by parts, the following consequences:
\begin{eqnarray}
\label{cycle2}
&&\int d^4x\, f\star g
=
\int d^4x\, f\cdot g = \int d^4x\, g\star f,
\\&&{}\label{cycle}
\int d^4x\, f_1\star f_2 \star\cdots\star f_N
=
\int d^4x\, f_2 \star\cdots\star f_N\star f_1.
\end{eqnarray}
Eq.~(\ref{cycle2}) leads us to conclude that the free part of an action in
noncommutative theory is the same as in the corresponding
commutative model, and from eq.~(\ref{cycle}) we see that
interaction terms which differ in the Lagrangian by a cyclic
permutation are the same in the action.
For example, in the theory of noncommutative comlex scalar field
theory which we shall study, there are two different interaction
terms \cite{0111208,0001215} and the action may be written as
\begin{eqnarray}
\nonumber
S&=&\int d^4x
 \Bigl(
   \partial_\mu\phi^*\star{}\partial^\mu\phi
   -m^2\phi^*\star{}\phi
\\&&\qquad{}
   -\frac{\lambda_a}{4!}\,\phi^*\star\phi\star\phi^*\star\phi
   -\frac{\lambda_b}{4!}\,\phi^*\star\phi^*\star\phi\star\phi
 \Bigr).
\label{action}
\end{eqnarray}
Here there are two possibilities of ordering the operators in
the interaction terms, therefore we can introduce
in general two independent coupling constants, in
contrast to the commutative case, where there is only one
interaction and only coupling constant.
This distinguishes the case of noncommutative complex scalar field
theory from the real one and makes the consideration of the
renormalization of composite operators in this theory also
interesting.

In the present paper we study this problem and compare it with
analogous problems both
in noncommutative real scalar field theory and in commutative
complex scalar field theory.

The paper is organized as follows.
In the next section we derive the classical energy-momentum
tensor of the noncommutative complex scalar field theory which
follows from the Noether's theorem and discuss some points
concerning its derivation in the noncommutative case.
In section~\ref{general} we present the general renormalization
structure of dimension four composite operators
and then in section~\ref{NCComOp} we renormalize these operators
in the one-loop approximation.
In section~\ref{EMT} we
find that the energy-momentum tensor is divergent and, in order
to make it finite, we need to add
"improving" terms to it.
These "improving" terms make the energy-momentum tensor
traceless (apart from being finite), but the latter is conserved in
the massless case only.
Also we study the renormalization of composite operators at zero
momentum transfer. This problem is considered in
section~\ref{NCComOp0}.
In section~\ref{EMV} we construct the energy-momentum vector of
the theory which follows from the Neother's procedure and find
it to be conserved and finite in the one-loop approximation.
In the summary we briefly discuss our results.

\section{Classical energy-momentum tensor}

In this section we define the energy-momentum tensor of the
noncommutative complex scalar field theory.
The action (\ref{action}) is invariant under the global
translation $x'^\mu=x^\mu+\varepsilon^\mu$,
$\varepsilon^\mu=const$
\begin{eqnarray}
\delta{}S
&=&
\int d^4x'\, {\cal L}'(x')
-
\int d^4x\, {\cal L}(x)
=0.
\label{deltaS}
\end{eqnarray}
The Lagrangian (as well as the field functions) is changed
under this transformation both because of changing the form of
the field functions\footnote{We have introduced the index $A$ at the
fields for the generality of analysis of the
energy-momentum tensor. For the theory under consideration we
put $\phi^1=\phi$, $\phi^2=\phi^*$.}
(the first line of (\ref{dS}))
$\bar\delta\phi^A(x)=\phi'^A(x)-\phi^A(x)=-\varepsilon^\mu\partial_\mu\phi^A$
and
because of changing the argument of the field functions
$\delta\phi^A(x)=\phi^A(x')-\phi^A(x)=\varepsilon^\mu\partial_\mu\phi^A$
(the second line of (\ref{dS})).
Therefore we can rewrite (\ref{deltaS}) as follows:
\begin{eqnarray}
\delta{}S
&=&
\int d^4x'\,\Bigl[ {\cal L}'(x')-{\cal L}(x') \Bigr]
\nonumber
\\
&&{}
+
\int d^4x'\, {\cal L}(x')
-
\int d^4x\, {\cal L}(x)=0.
\label{dS}
\end{eqnarray}
Since $d^4x'=d^4x$, the last line of (\ref{dS}) to the first
order in $\varepsilon^\mu$ reads
\begin{eqnarray}
\int d^4x\,\Bigl[ {\cal L}(x+\varepsilon)-{\cal L}(x) \Bigr]
=
\int d^4x\, \varepsilon^\mu\partial_\mu{\cal{}L}.
\label{argum}
\end{eqnarray}

As far as the first line of (\ref{dS}) is concerned, we transform
it as follows (changing $x'\to{}x$):
\begin{eqnarray}
\bar\delta{}S
&=&
\int d^4x\, \Bigl[{\cal L}'(x)-{\cal L}(x)\Bigr]
\nonumber
\\
&=&
\int d^4x\, \Bigl[
\frac{\partial{\cal L}}{\partial\phi^A}\bar\delta\phi^A
+
\frac{\partial{\cal
L}}{\partial\phi,_{\mu}^A}\bar\delta\phi,_{\mu}^A
\Bigr].
\end{eqnarray}
It should be noted here that in calculating this expression (and
the equation of motion in the following) we have used the cyclic
property (\ref{cycle})
and therefore the integration region in the noncommutative
directions is not arbitrary.
Since
\begin{math}
\bar\delta\phi,_{\mu}^A
=
\phi',_{\mu}^A-\phi,_{\mu}^A
=
(\phi'^A-\phi^A),_{\mu}
=
(\bar\delta\phi^A),_{\mu}
\end{math}
and using the equations of motion we have
\begin{eqnarray}
\bar\delta{}S
&=&
\int d^4x \Bigl[
\partial_\mu\frac{\partial{\cal{}L}}{\partial\phi,^A_{\mu}}
\,\bar\delta\phi^A
+
\frac{\partial{\cal{}L}}{\partial\phi,^A_{\mu}}\,
\partial_\mu\bar\delta\phi^A
\Bigr]
\nonumber
\\
&=&
\int d^4x
\partial_\mu
\Bigl(
\frac{\partial{\cal{}L}}{\partial\phi,^A_{\mu}}
\,\bar\delta\phi^A
\Bigr)
=
-
\int d^4x\,\varepsilon^\nu
\partial_\mu
\Bigl(
\frac{\partial{\cal{}L}}{\partial\phi,^A_{\mu}}
\,\partial_{\nu}\phi^A
\Bigr).
\label{form}
\end{eqnarray}

Collecting together (\ref{argum}), (\ref{form}) and using the
arbitrariness of $\varepsilon^\nu$ we find that
\begin{eqnarray}
\int d^4x \partial_\mu
\Bigl[
\frac{\partial{\cal{}L}}{\partial\phi^{A,\mu}}
\,\partial_{\nu}\phi^A
-
\eta_{\mu\nu}{\cal{}L}
\Bigr]
=0.
\label{GenT}
\end{eqnarray}
In the case of spatial noncommutativity $\theta^{0i}=0$ there
are no time derivatives in the star product (\ref{theta})
and
therefore the properties (\ref{cycle2}), (\ref{cycle}) still
hold when the integration is performed in space coordinates only.
Therefore in eq.~(\ref{GenT}) the time integration region is
arbitrary and we can write that
\begin{eqnarray}
\int d^3x \partial_\mu
\Bigl[
\frac{\partial{\cal{}L}}{\partial\phi^{A,\mu}}
\,\partial_{\nu}\phi^A
-
\eta_{\mu\nu}{\cal{}L}
\Bigr]
=0
\label{GenT3}
\end{eqnarray}
and so, the quantity
\begin{eqnarray}
\int d^3x
\Bigl[
\frac{\partial{\cal{}L}}{\partial\phi^{A,0}}
\,\partial_{\nu}\phi^A
-
\eta_{0\nu}{\cal{}L}
\Bigr]
\end{eqnarray}
is conserved.
As a result, in noncommutative field theories there are only
global conservation laws of the energy-momentum and only in the case
of spatial noncommutativity $\theta^{0i}=0$.

Substituting the Lagrangian of the theory under consideration
(\ref{action}) in (\ref{GenT3}) we find
\begin{eqnarray}
&&
\label{Noether}
\int\partial^\mu \tilde{T}_{\mu\nu}\, d^3x=0,
\\&&
\qquad
{}
\nonumber
\tilde{T}_{\mu\nu}=
   \partial_\mu\phi^*\star\partial_\nu\phi
  +\partial_\nu\phi^*\star\partial_\mu\phi
  -\eta_{\mu\nu}
     \biggl(
          \partial_\alpha\phi^*\star{}\partial^\alpha\phi
         -m^2\phi^*\star{}\phi
\\&&\qquad\qquad{}
         -\frac{\lambda_a}{4!}\,\phi^*\star\phi\star\phi^*\star\phi
         -\frac{\lambda_b}{4!}\,\phi^*\star\phi^*\star\phi\star\phi
     \biggr).
\label{Noether2}
\end{eqnarray}
The expression (\ref{Noether}) is the basis for defining the
energy-momentum tensor of the theory.
Since the properties (\ref{cycle2}), (\ref{cycle}) are still
valid, we must consider
all possibilities of field ordering in the energy-momentum
tensor which lead us to
(\ref{Noether2}).
As a result,
in contrast to the commutative case, the energy-momentum tensor
of the noncommutative complex scalar field theory cannot be
defined unambigously and we may write it only as follows:
\begin{eqnarray}
T_{\mu\nu}
&=&
c_1\biggl(
   \partial_\mu\phi^*\star\partial_\nu\phi
  +\partial_\nu\phi^*\star\partial_\mu\phi
 \biggr)
+
(1-c_1)
   \biggl(
   \partial_\mu\phi\star\partial_\nu\phi^*
  +\partial_\nu\phi\star\partial_\mu\phi^*
   \biggr)
\nonumber
\\&&{}
-\eta_{\mu\nu}\biggl(
   c_2\,\partial_\alpha\phi^*\star\partial^\alpha\phi
   +(1-c_2)\,\partial_\alpha\phi\star\partial^\alpha\phi^*
   \biggr)
\nonumber
\\&&{}
+\eta_{\mu\nu}\,m^2
   \biggl(
   c_3\,\phi^*\star\phi
   +(1-c_3)\,\phi\star\phi^*
      \biggr)
\nonumber
\\&&{}
+\eta_{\mu\nu}\,\frac{\lambda_a}{4!}\,
\biggl(
c_4\, \phi^*\star\phi\star\phi^*\star\phi
+(1-c_4)\, \phi\star\phi^*\star\phi\star\phi^*
\biggr)
\nonumber
\\&&{}
+\eta_{\mu\nu}\,\frac{\lambda_b}{4!}\, c_5\, \phi^*\star\phi^*\star\phi\star\phi
+\eta_{\mu\nu}\,\frac{\lambda_b}{4!}\, c_6\, \phi\star\phi\star\phi^*\star\phi^*
\nonumber
\\&&{}
+\eta_{\mu\nu}\,\frac{\lambda_b}{4!}\,\frac{1-c_5-c_6}{2} \biggl(\phi^*\star\phi\star\phi\star\phi^* + \phi\star\phi^*\star\phi^*\star\phi\biggr),
\label{emt}
\end{eqnarray}
with $c_1$, \ldots, $c_6$ being arbitrary real numbers, which
define all possibilities of field ordering.
Also we have demanded here that the energy-momentum tensor
(\ref{emt}) be symmetric and real.
If we substitute (\ref{emt}) in (\ref{Noether}) we find that all
the arbitrary constant are cancelled.

As a result, we see that the energy-momentum tensor of the
noncommutative complex scalar field theory is defined up to six
arbitrary real constants.
This separates the theory under consideration both from the
corresponding commutative theory and from the noncommutative
theory of real scalar field, where the energy-momentum tensors
are defined unambigously.

\section{General renormalization structure of dimension four
composite operators}\label{general}

In order to construct the renormalized energy-momentum tensor,
we have to renormalize all the composite operators which enter
into it. Before starting an explicit calculation of the renormalization of
composite operators, let us describe the general situation.
As well known (see e.g. \cite{Collins}), in order to renormalize some composite
operator we need, in general, to take into account all composite
operators with the same mass dimension, tensor structure, and
symmetries. One can show that it is sufficient to renormalize the
operators which constitute a basis of such operators \cite{Collins}.

The operators which enter into the energy-momentum tensor
(\ref{emt}) are hermitian operators with mass dimension four.
Let us construct a corresponding operator basis.
From a general point of view it is convenient also to include into the
basis, together with other operators, the operators which are not
renormalized and are related to the field equations in terms of bare (with
index 0) or renormalized quantities\footnote{ We use dimensional
regularization and keep the renormalized coupling constants $\lambda_a$
and $\lambda_b$ to be dimensionless, therefore we introduce by standard
way an arbitrary parameter $\mu$ of mass dimension.}
\begin{eqnarray}
L_0 &=& (\partial^2+m_0^2)\phi_0
+2\frac{\lambda_{a0}}{4!}\phi_0\star\phi_0^*\star\phi_0
+\frac{\lambda_{b0}}{4!}\phi_0^*\star\phi_0\star\phi_0
+\frac{\lambda_{b0}}{4!}\phi_0\star\phi_0\star\phi_0^*,
\\
L&=&
(\partial^2+m^2)\phi
+2\frac{\mu^{4-d}\lambda_{a}}{4!}\phi\star\phi^*\star\phi
+\frac{\mu^{4-d}\lambda_{b}}{4!}\phi^*\star\phi\star\phi
+\frac{\mu^{4-d}\lambda_{b}}{4!}\phi\star\phi\star\phi^*.
\end{eqnarray}
We construct the basis of the scalar hermitian composite
operators with mass dimension four as follows:
\begin{eqnarray}
Q_0=\left(
\begin{array}{c}
\phi_0^*\star\phi_0\star\phi_0^*\star\phi_0\\
\phi_0\star\phi_0^*\star\phi_0\star\phi_0^*\\
\phi_0^*\star\phi_0^*\star\phi_0\star\phi_0\\
\phi_0\star\phi_0\star\phi_0^*\star\phi_0^*\\
\phi_0^*\star\phi_0\star\phi_0\star\phi_0^*+\\
\hspace{2em}+\phi_0\star\phi_0^*\star\phi_0^*\star\phi_0\\
m_0^2\,\phi_0^*\star\phi_0\\
m_0^2\,\phi_0\star\phi_0^*\\
\partial^2(\phi_0^*\star\phi_0)\\
\partial^2(\phi_0\star\phi_0^*)\\
\phi_0^*\star{}L_0+L_0^*\star\phi_0\\
\phi_0\star{}L_0^*+L_0\star\phi_0^*
\end{array}
\right)
&\mbox{and}&
[Q]=\left(
\begin{array}{c}
[\phi^*\star\phi\star\phi^*\star\phi]\\{}
[\phi\star\phi^*\star\phi\star\phi^*]\\{}
[\phi^*\star\phi^*\star\phi\star\phi]\\{}
[\phi\star\phi\star\phi^*\star\phi^*]\\
\bigl[
\phi^*\star\phi\star\phi\star\phi^*+\\
\hspace{2em}+\phi\star\phi^*\star\phi^*\star\phi
\bigr]\\
m^2\,[\phi^*\star\phi]\\
m^2\,[\phi\star\phi^*]\\
\partial^2[\phi^*\star\phi]\\
\partial^2[\phi\star\phi^*]\\{}
[\phi^*\star{}L+L^*\star\phi]\\{}
[\phi\star{}L^*+L\star\phi^*]
\end{array}
\right),
\label{bases}
\end{eqnarray}
for the bare and the renormalized operators respectively.
Due to the noncommutativity of the multiplication rule (\ref{theta}), the
operator basis (\ref{bases}) contains more operators in comparison
with the corresponding commutative theory because we may order the
fields in composite operators in different ways.  For example, the
hermitian operator $(\phi^*\phi)^2$ in the commutative theory is a
prototype for five different hermitian operators in the noncommutative
theory (\ref{bases}).  This situation is typical for any noncommutative
field theory.  These operator basis are mixed by renormalization.
By dimensional and symmetry analysis it may be shown that no more
operators are required.  We will write the relation (\ref{bases}) in
the form
\begin{eqnarray}
&& Q_0=Z[Q], \label{Z}
\end{eqnarray}
with
$Z$ being a mixing matrix.  In the next section we calculate this
mixing matrix $Z$ and some more renormalization relations for composite
operators, which we will use for the renormalization of the
energy-momentum tensor in the one-loop approximation.

\section{One-loop renormalization of dimension four composite operators}\label{NCComOp}

In this section we carry out the one-loop renormalization
of all composite operators
entering into the expression for the energy-momentum tensor
(\ref{emt}) of the noncommutative complex scalar field theory.
The general procedure of constructing the renormalized operators
which is valid for both commutative and noncommutative theories
was described in \cite{0303186}.
Here we follow this procedure.
Before starting to renormalize the composite operators we need
to renormalize all parameters of the theory.
The one-loop renormalization of the field, the mass, and the
coupling constants of the model may be found as a particular
case of \cite{0207086} and reads
\begin{eqnarray}
\phi_0&=&\phi,
\\
m^2_0&=&
\left(
 1-\frac{1}{(d-4)(4\pi)^2}\frac{2\lambda_a+\lambda_b}{3!}
\right)m^2,
\\
\frac{\lambda_{a0}}{4!}
&=&
\mu^{4-d}\frac{\lambda_a}{4!}
-\frac{2\mu^{4-d}}{(d-4)(4\pi)^2}
\frac{4\lambda_a^2+\lambda_b^2}{(4!)^2},
\\
\frac{\lambda_{b0}}{4!}
&=&
\mu^{4-d}\frac{\lambda_b}{4!}
-\frac{2\mu^{4-d}}{(d-4)(4\pi)^2}
\frac{\lambda_b}{4!}\,
\frac{4\lambda_a+\lambda_b}{4!}.
\end{eqnarray}

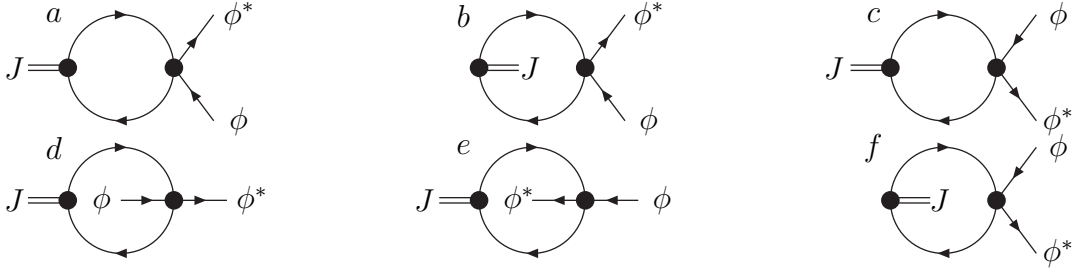
\begin{figure}[t]
\begin{picture}(450,100)(0,0)
\Vertex(50,70){3.5}
\Vertex(90,70){3.5}
\ArrowArcn(70,70)(20,0,180)
\ArrowArcn(70,70)(20,180,360)
\Line(35,69)(50,69)
\Line(35,71)(50,71)
\ArrowLine(90,70)(105,90)
\ArrowLine(105,50)(90,70)
\Text(30,70)[c]{\mbox{}$J$}
\Text(115,90)[c]{\mbox{}$\phi^*$}
\Text(115,50)[c]{\mbox{}$\phi$}
\Text(45,90)[c]{$a$}
\Vertex(205,70){3.5}
\Vertex(245,70){3.5}
\ArrowArcn(225,70)(20,0,180)
\ArrowArcn(225,70)(20,180,360)
\Line(205,69)(220,69)
\Line(205,71)(220,71)
\ArrowLine(245,70)(260,90)
\ArrowLine(260,50)(245,70)
\Text(225,70)[c]{\mbox{}$J$}
\Text(270,90)[c]{\mbox{}$\phi^*$}
\Text(270,50)[c]{\mbox{}$\phi$}
\Text(200,90)[c]{$b$}
\Vertex(360,70){3.5}
\Vertex(400,70){3.5}
\ArrowArcn(380,70)(20,0,180)
\ArrowArcn(380,70)(20,180,360)
\Line(345,69)(360,69)
\Line(345,71)(360,71)
\ArrowLine(415,90)(400,70)
\ArrowLine(400,70)(415,50)
\Text(340,70)[c]{\mbox{}$J$}
\Text(425,90)[c]{\mbox{}$\phi$}
\Text(425,50)[c]{\mbox{}$\phi^*$}
\Text(355,90)[c]{$c$}
\Vertex(50,20){3.5}
\Vertex(90,20){3.5}
\ArrowArcn(70,20)(20,0,180)
\ArrowArcn(70,20)(20,180,360)
\Line(35,19)(50,19)
\Line(35,21)(50,21)
\ArrowLine(70,20)(90,20)
\ArrowLine(90,20)(110,20)
\Text(30,20)[c]{\mbox{}$J$}
\Text(120,20)[c]{\mbox{}$\phi^*$}
\Text(63,20)[c]{\mbox{}$\phi$}
\Text(45,40)[c]{$d$}
\Vertex(205,20){3.5}
\Vertex(245,20){3.5}
\ArrowArcn(225,20)(20,0,180)
\ArrowArcn(225,20)(20,180,360)
\Line(205,19)(190,19)
\Line(205,21)(190,21)
\ArrowLine(265,20)(245,20)
\ArrowLine(245,20)(225,20)
\Text(185,20)[c]{\mbox{}$J$}
\Text(215,20)[l]{\mbox{}$\phi^*$}
\Text(275,20)[c]{\mbox{}$\phi$}
\Text(200,40)[c]{$e$}
\Vertex(360,20){3.5}
\Vertex(400,20){3.5}
\ArrowArcn(380,20)(20,0,180)
\ArrowArcn(380,20)(20,180,360)
\Line(360,19)(375,19)
\Line(360,21)(375,21)
\ArrowLine(415,40)(400,20)
\ArrowLine(400,20)(415,0)
\Text(380,20)[c]{\mbox{}$J$}
\Text(425,40)[c]{\mbox{}$\phi$}
\Text(425,0)[c]{\mbox{}$\phi^*$}
\Text(355,40)[c]{$f$}
\end{picture}
\caption{Divergent diagrams corresponding to the operator
$m^2\,\phi^*\star\phi$}\label{f*f}
\end{figure}

Let us consider the renormalization of the composite operator
$m^2\,\phi^*\star\phi$.
Following the procedure described in \cite{0303186}, in order to
renormalize this operator we should add to the action
(\ref{action}) the term
\begin{eqnarray}
\label{Jff}
m^2\,\int d^dx\, J\star\phi^*\star\phi,
\end{eqnarray}
with $J$ being some arbitrary function (source),
and then calculate all divergent one particle irreducible diagrams
linear in $J$. There are six types of such diagrams which are
shown in Figure~\ref{f*f}.
Performing the Fourier transform of
the fields $\phi^*$, $\phi$ and the source $J$
\begin{eqnarray}
&&
\phi^*(x)=\int \left(\frac{dp_1}{2\pi}\right)^{\!d}\, e^{ip_1x}\, \tilde{\phi}_*(p_1)
\equiv\int_{p_1} e^{ip_1x}\, \tilde{\phi}_*(p_1),
\\&&
\phi(x)=\int_{p_2} e^{ip_2x}\, \tilde{\phi}(p_2),
\\&&
J(x)=\int_{k}  e^{ikx}\, \tilde{J}(k),
\end{eqnarray}
we get the following expression in momentum space for the first two diagrams (which correspond to the
$\lambda_a$ interaction term) in Figure~\ref{f*f}
\begin{eqnarray}
&&
2i\,\frac{\lambda_a}{4!}\, m^2
\int_{kp_1p_2}(2\pi)^d\delta(k+p_1+p_2)\tilde{J}(k)\tilde{\phi}_*(p_1)\tilde{\phi}(p_2)\times
\nonumber
\\&&\qquad\times
\biggl\{
e^{-\frac{i}{2}p_1\theta{}p_2}
\int_{p}
\frac{1}{(p^2-m^2)((p+k)^2-m^2)}
\label{f*fa}
\\&&\quad\qquad{}
+
e^{-\frac{i}{2}p_2\theta{}p_1}
\int_{p}
\frac{e^{ip\theta{}k}}{(p^2-m^2)((p+k)^2-m^2)}
\,
\biggr\}.
\label{f*fb}
\end{eqnarray}
The expression (\ref{f*fa}) corresponds to diagram $a$ in
Figure~\ref{f*f} and has a UV divergence at any external momenta
$k$, $p_1$, $p_2$ (the so-called "planar" diagram). The
expression (\ref{f*fb}) corresponds to diagram $b$ (so-called
"non-planar" diagram) and displays UV/IR mixing \cite{9912072}:
its divergence depends on the value of $\theta^{\mu\nu}k_\nu$.
If $\theta^{\mu\nu}k_\nu=0$, then we have a UV divergence as in
the commutative theory.
If $\theta^{\mu\nu}k_\nu\neq{}0$ then the integral (\ref{f*fb}) is
UV finite, but if we were to put $k_\nu=0$ after carrying out the
integration, then we would find it to be divergent (IR
divergence).
We suppose that
$\theta^{\mu\nu}k_\nu\neq{}0$, so only the expression (\ref{f*fa})
contains a UV divergence and for its subtraction we need to add a
counterterm in the effective action. An analogous consideration is
valid for the other four diagrams in Figure~\ref{f*f}. We have
for them the following expressions in momentum space:
\begin{eqnarray}
&&
i\,\frac{\lambda_b}{4!}\, m^2
\int_{kp_1p_2}(2\pi)^d\delta(k+p_1+p_2)\tilde{J}(k)\tilde{\phi}_*(p_1)\tilde{\phi}(p_2)\times
\nonumber
\\&&\qquad\times
\biggl\{
e^{-\frac{i}{2}p_2\theta{}p_1}
\int_{p}
\frac{1}{(p^2-m^2)((p+k)^2-m^2)}
\label{f*fc}
\\&&\quad\qquad{}
+
e^{-\frac{i}{2}p_2\theta{}p_1}
\int_{p}
\frac{e^{-ip\theta{}p_2}}{(p^2-m^2)((p+k)^2-m^2)}
\label{f*fd}
\\&&\quad\qquad{}
+
e^{-\frac{i}{2}p_1\theta{}p_2}
\int_{p}
\frac{e^{-ip\theta{}p_1}}{(p^2-m^2)((p+k)^2-m^2)}
\label{f*fe}
\\&&\quad\qquad{}
+
e^{-\frac{i}{2}p_1\theta{}p_2}
\int_{p}
\frac{e^{ip\theta{}k}}{(p^2-m^2)((p+k)^2-m^2)}
\,
\biggr\}.
\label{f*ff}
\end{eqnarray}
The expression (\ref{f*fc}) corresponds to diagram $c$ (planar
diagram) and is UV divergent at any external momenta.
Expressions (\ref{f*fd}), (\ref{f*fe}), (\ref{f*ff}) correspond
to diagrams $d$, $e$, $f$ in the figure.
These diagrams are non-planar and their divergences depend on
values of $\theta^{\mu\nu}p_{2\nu}$, $\theta^{\mu\nu}p_{1\nu}$,
$\theta^{\mu\nu}k_\nu$ respectively.
This situation is analogous to that when we consider diagram $b$
in Figure~\ref{f*f}.
As in that case, we suppose that
$\theta^{\mu\nu}p_{2\nu}\neq{}0$,
$\theta^{\mu\nu}p_{1\nu}\neq{}0$, $\theta^{\mu\nu}k_\nu\neq{}0$,
so these diagrams have no UV divergences.
Using dimensional regularization we find the UV divergent parts
of (\ref{f*fa}) and (\ref{f*fc})
\begin{eqnarray}
&&\frac{1}{(d-4)(4\pi)^2}\,\frac{\lambda_a}{3!}\,\,m^2
\int d^4x\,J\star\phi^*\star\phi
\\
&&\frac{1}{(d-4)(4\pi)^2}\,\frac{2\lambda_b}{4!}\,\,m^2
\int d^4x\,J\star\phi\star\phi^*
\end{eqnarray}
respectively.
As a result we get the following expression connecting the bare
and renormalized operators:
\begin{eqnarray}
m^2_0\,\phi_0^*\star\phi_0
&=&
\left(
       1-\frac{1}{(d-4)(4\pi)^2}\frac{\lambda_a+\lambda_b}{3!}
\right)
m^2\,[\phi^*\star\phi]
\nonumber{}
\\&&{}
+
\frac{1}{(d-4)(4\pi)^2}\frac{2\lambda_b}{4!}\,m^2[\phi\star\phi^*].
\label{first+}
\end{eqnarray}
From (\ref{first+}) we see that there is operator mixing here:
in order to renormalize the operator $m_0^2\,\phi_0^*\star\phi_0$, we
have to take into account the operator $m^2\,[\phi\star\phi^*]$
besides $m^2\,[\phi^*\star\phi]$ which are different due to
noncommutativity of the multiplication (\ref{theta}).
Also the expression (\ref{first+}) differs from the corresponding
renormalization relation in the commutative theory where, as it may
easily be shown, one has
\begin{eqnarray}
m_0^2\,\phi_0^*\phi_0
&=&
m^2\,[\phi^*\phi].
\label{comff}
\end{eqnarray}
This situation is similar to that in the noncommutative real
scalar field theory: in both cases the
corresponding renormalization relations in noncommutative and
commutative theories have different form.

In complete analogy we may calculate the renormalization of the
operator $m_0^2\,\phi_0\star\phi_0^*$ which differs from the
previously renormalized operator by the following exchange of
the fields: $\phi_0\leftrightarrow\phi_0^*$. The action
(\ref{action}) also has this symmetry, so we may find the
renormalization relation for the operator
$m_0^2\,\phi_0\star\phi_0^*$ by exchanging
$\phi_0\leftrightarrow\phi_0^*$
(and $\phi\leftrightarrow\phi^*$)
in (\ref{first+}).
As a result we have
\begin{eqnarray}
m^2_0\phi_0\star\phi_0^*
&=&
\left(
       1-\frac{1}{(d-4)(4\pi)^2}\frac{\lambda_a+\lambda_b}{3!}
\right)
m^2[\phi\star\phi^*]
\nonumber
\\&&{}
+
\frac{1}{(d-4)(4\pi)^2}\frac{2\lambda_b}{4!}\,m^2[\phi^*\star\phi].
\label{first}
\end{eqnarray}
In the following we shall ignore the order of lines in vertices,
which in noncommutative theories commonly are depicted in the
proper cyclic order, reflecting the order of the fields in the
action and the property (\ref{cycle}).
Therefore, for example, we could have shown all diagrams represented in
Figure~\ref{f*f} with the help of only one
of them, usually $a$ or $c$.

\begin{figure}[t]
\begin{picture}(450,55)(0,-5)
\Vertex(130,20){3.5}
\Vertex(170,20){3.5}
\ArrowArcn(150,20)(20,0,180)
\ArrowArcn(150,20)(20,180,360)
\Line(110,19)(130,19)
\Line(110,21)(130,21)
\ArrowLine(170,20)(190,40)
\ArrowLine(190,0)(170,20)
\Text(100,20)[c]{\mbox{}$J^{\mu\nu}$}
\Text(199,40)[c]{\mbox{}$\phi^*$}
\Text(197,0)[c]{\mbox{}$\phi$}
\Text(125,40)[c]{$a$}
\Vertex(280,20){3.5}
\Vertex(310,37){3.5}
\Vertex(310,3){3.5}
\ArrowArcn(300,20)(20,60,300)
\ArrowArcn(300,20)(20,180,60)
\ArrowArcn(300,20)(20,300,180)
\Line(260,19)(280,19)
\Line(260,21)(280,21)
  \Text(250,20)[c]{\mbox{}$J^{\mu\nu}$}
\ArrowLine(320,50)(310,37)
  \Text(327,50)[c]{\mbox{}$\phi$}
\ArrowLine(310,37)(340,45)
  \Text(350,45)[c]{\mbox{}$\phi^*$}
\ArrowLine(320,-10)(310,3)
  \Text(327,-10)[c]{\mbox{}$\phi$}
\ArrowLine(310,3)(340,-5)
  \Text(350,-5)[c]{\mbox{}$\phi^*$}
\Text(275,40)[c]{$b$}
\end{picture}
\caption{Divergent diagrams corresponding to the operator
$\partial_\mu\phi^*\star\partial_\nu\phi$}\label{df*df}
\end{figure}
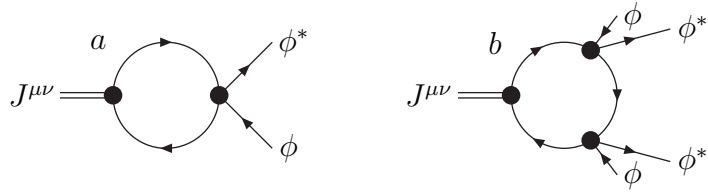

Now let us consider the renormalization of the operator
$\partial_\mu\phi^*\star\partial_\nu\phi$.
According the procedure described in \cite{0303186} we should
add to the action (\ref{action}) the term
\begin{eqnarray}
\int d^dx\, J^{\mu\nu}\star
\partial_\mu\phi^*\star\partial_\nu\phi
\end{eqnarray}
and renormalize all divergent one particle irreducible diagrams linear in
$J^{\mu\nu}$.
These diagrams are shown in Figure~\ref{df*df}.
As was explained above these diagrams are concise and contain
both vertices with $\lambda_a$ and $\lambda_b$ interactions and
both planar and nonplanar contributions to the generating
functional of one particle irreducible diagrams.
As usual
we suppose that external momenta are not zero, therefore only
planar diagrams have UV divergences and their contribution to
the effective action is
\begin{eqnarray}
&&
i\mu^{4-d}
\int_{kp_1p_2}(2\pi)^d\delta(k+p_1+p_2)
\tilde{J}{}^{\mu\nu}(k)
\tilde{\phi}_*(p_1)
\tilde{\phi}(p_2)
\Biggl(
2\frac{\lambda_a}{4!}\,
e^{-\frac{i}{2}p_1\theta{}p_2}
+
\frac{\lambda_b}{4!}\,
e^{\frac{i}{2}p_1\theta{}p_2}
\Biggr)
\times
\nonumber
\\&&\qquad\qquad{}
\times
\int_p\frac{(p_\mu-k_\mu)p_\nu}{(p^2-m^2)((p-k)^2-m^2)}
\label{df*dfa}
\end{eqnarray}
for diagram $a$ and
\begin{eqnarray}
&&
i\mu^{8-2d}
\int_{kp_1p_2p_3p_4}(2\pi)^d\delta(k+p_1+p_2+p_3+p_4)
\tilde{J}{}^{\mu\nu}(k)
\tilde{\phi}_*(p_1)
\tilde{\phi}(p_2)
\tilde{\phi}_*(p_3)
\tilde{\phi}(p_4)
\times
\nonumber
\\
&&\qquad\qquad{}
\times
\Biggl\{
4\left(\frac{\lambda_a}{4!}\right)^2
e^{-\frac{i}{2}p_1\theta{}(p_2+p_3+p_4)}
e^{-\frac{i}{2}p_2\theta{}(p_3+p_4)}
e^{-\frac{i}{2}p_3\theta{}p_4}
+
\nonumber
\\
&&\qquad\qquad\qquad
+
\left(\frac{\lambda_b}{4!}\right)^2
e^{-\frac{i}{2}p_2\theta{}(p_1+p_3+p_4)}
e^{-\frac{i}{2}p_1\theta{}(p_3+p_4)}
e^{-\frac{i}{2}p_4\theta{}p_3}
+
\nonumber
\\
&&\qquad\qquad\qquad
+
2
\frac{\lambda_a}{4!}
\frac{\lambda_b}{4!}
e^{-\frac{i}{2}p_2\theta{}(p_1+p_3+p_4)}
e^{-\frac{i}{2}p_1\theta{}(p_3+p_4)}
e^{-\frac{i}{2}p_3\theta{}p_4}
+
\nonumber
\\
&&\qquad\qquad\qquad
+
2
\frac{\lambda_a}{4!}
\frac{\lambda_b}{4!}
e^{-\frac{i}{2}p_1\theta{}(p_2+p_3+p_4)}
e^{-\frac{i}{2}p_2\theta{}(p_3+p_4)}
e^{-\frac{i}{2}p_4\theta{}p_3}
\Biggr\}
\times
\nonumber
\\
&&\qquad{}
\times
\int_{p}
\frac{(p_\mu-k_\mu)p_\nu}{(p^2-m^2)((p+p_3+p_4)^2-m^2)((p-k)^2-m^2)}
\nonumber
\\
&+&
i\mu^{8-2d}
\left(\frac{\lambda_b}{4!}\right)^2
\int_{kp_1p_2p_3p_4}\tilde{\delta}(k+p_1+p_2+p_3+p_4)
\tilde{J}{}^{\mu\nu}(k)
\tilde{\phi}_*(p_1)
\tilde{\phi}(p_2)
\tilde{\phi}_*(p_3)
\tilde{\phi}(p_4)
\times
\nonumber
\\
&&\qquad\qquad\qquad\qquad\qquad{}
\times
e^{-\frac{i}{2}p_1\theta{}(p_2+p_3+p_4)}
e^{-\frac{i}{2}p_3\theta{}(p_2+p_4)}
e^{-\frac{i}{2}p_2\theta{}p_4}
\times
\nonumber
\\
&&\qquad{}
\times
\int_{p}
\frac{(p_\mu-k_\mu)p_\nu}{(p^2-m^2)((p+p_2+p_4)^2-m^2)((p-k)^2-m^2)}
\label{df*dfb}
\end{eqnarray}
for diagram $b$. From (\ref{df*dfa}) and (\ref{df*dfb}) UV
divergences may easily be extracted and the result is
\begin{eqnarray}
&&
\frac{2}{(d-4)(4\pi)^2}
\frac{\lambda_a}{4!}
\int d^4x\, \phi^*\star\phi\star
\biggl[
\frac{1}{6}\eta_{\mu\nu}\partial^2
+
m^2\eta_{\mu\nu}
+
\frac{1}{3}\partial^2_{\mu\nu}
\biggr]
J^{\mu\nu}
\nonumber
\\
&+&
\frac{1}{(d-4)(4\pi)^2}
\frac{\lambda_b}{4!}
\int d^4x\, \phi\star\phi^*\star
\biggl[
\frac{1}{6}\eta_{\mu\nu}\partial^2
+
m^2\eta_{\mu\nu}
+
\frac{1}{3}\partial^2_{\mu\nu}
\biggr]
J^{\mu\nu}
\label{df*dfa2}
\end{eqnarray}
for expression (\ref{df*dfa}) and
\begin{eqnarray}
&&
\frac{2\mu^{4-d}}{(d-4)(4\pi)^2}
\left(\frac{\lambda_a}{4!}\right)^2
\eta_{\mu\nu}
\int d^4x\, J^{\mu\nu}\star\phi^*\star\phi\star\phi^*\star\phi
\nonumber
\\
&+&
\frac{1/2\mu^{4-d}}{(d-4)(4\pi)^2}
\left(\frac{\lambda_b}{4!}\right)^2
\eta_{\mu\nu}
\int d^4x\, J^{\mu\nu}\star\phi\star\phi^*\star\phi\star\phi^*
\nonumber
\\
&+&
\frac{\mu^{4-d}}{(d-4)(4\pi)^2}\,
\frac{\lambda_a}{4!}\,
\frac{\lambda_b}{4!}\,
\eta_{\mu\nu}
\int d^4x\, J^{\mu\nu}\star\phi\star\phi^*\star\phi^*\star\phi
\nonumber
\\
&+&
\frac{\mu^{4-d}}{(d-4)(4\pi)^2}\,
\frac{\lambda_a}{4!}\,
\frac{\lambda_b}{4!}\,
\eta_{\mu\nu}
\int d^4x\, J^{\mu\nu}\star\phi^*\star\phi\star\phi\star\phi^*
\nonumber
\\
&+&
\frac{1/2\mu^{4-d}}{(d-4)(4\pi)^2}
\left(\frac{\lambda_b}{4!}\right)^2
\eta_{\mu\nu}
\int d^4x\, J^{\mu\nu}\star\phi^*\star\phi^*\star\phi\star\phi
\label{df*dfb2}
\end{eqnarray}
for expression (\ref{df*dfb}).
Using (\ref{df*dfa2}) and (\ref{df*dfb2}) we get the one-loop
renormalization relation for the operator
$\partial_\mu\phi^*\star\partial_\nu\phi$
\begin{eqnarray}
\partial_\mu\phi_0^*\star\partial_\nu\phi_0
&=&
[\partial_\mu\phi^*\star\partial_\nu\phi]
\nonumber{}
\\&&
\nonumber{}
+\frac{2}{(d-4)(4\pi)^2}\frac{\lambda_a}{4!}
\left(\frac{1}{6}\eta_{\mu\nu}\partial^2
     +\frac{1}{3}\partial^2_{\mu\nu}
     +\eta_{\mu\nu}\,m^2
\right)
[\phi^*\star\phi]
\\&&
\nonumber{}
+\frac{1}{(d-4)(4\pi)^2}\frac{\lambda_b}{4!}
\left(\frac{1}{6}\eta_{\mu\nu}\partial^2
     +\frac{1}{3}\partial^2_{\mu\nu}
     +\eta_{\mu\nu}\,m^2
\right)
[\phi\star\phi^*]
\\&&
\nonumber{}
+\frac{2\mu^{4-d}}{(d-4)(4\pi)^2}\left(\frac{\lambda_a}{4!}\right)^2
\eta_{\mu\nu}[\phi^*\star\phi\star\phi^*\star\phi]
\\&&
\hspace{-15mm}
\nonumber{}
+\frac{1/2\mu^{4-d}}{(d-4)(4\pi)^2}\left(\frac{\lambda_b}{4!}\right)^2
\eta_{\mu\nu}
\Bigl(
 [\phi\star\phi^*\star\phi\star\phi^*]
 +[\phi^*\star\phi^*\star\phi\star\phi]
\Bigr)
\\&&
\hspace{-15mm}
{}
+\frac{\mu^{4-d}}{(d-4)(4\pi)^2}\frac{\lambda_a}{4!}\frac{\lambda_b}{4!}
\eta_{\mu\nu}
\Bigl(
 [\phi\star\phi^*\star\phi^*\star\phi]
 +[\phi^*\star\phi\star\phi\star\phi^*]
\Bigr).
\label{df*df=}
\end{eqnarray}

The renormalization of the operator
$\partial_\mu\phi\star\partial_\nu\phi^*$ may be found by
exchanging $\phi\leftrightarrow\phi^*$ in (\ref{df*df=})
\begin{eqnarray}
\partial_\mu\phi_0\star\partial_\nu\phi_0^*
&=&
[\partial_\mu\phi\star\partial_\nu\phi^*]
\nonumber
\\&&
\nonumber{}
+\frac{2}{(d-4)(4\pi)^2}\frac{\lambda_a}{4!}
\left(\frac{1}{6}\eta_{\mu\nu}\partial^2
     +\frac{1}{3}\partial^2_{\mu\nu}
     +\eta_{\mu\nu}\,m^2
\right)
[\phi\star\phi^*]
\\&&
\nonumber{}
+\frac{1}{(d-4)(4\pi)^2}\frac{\lambda_b}{4!}
\left(\frac{1}{6}\eta_{\mu\nu}\partial^2
     +\frac{1}{3}\partial^2_{\mu\nu}
     +\eta_{\mu\nu}\,m^2
\right)
[\phi^*\star\phi]
\\&&
\nonumber{}
+\frac{2\mu^{4-d}}{(d-4)(4\pi)^2}\left(\frac{\lambda_a}{4!}\right)^2
\eta_{\mu\nu}[\phi\star\phi^*\star\phi\star\phi^*]
\\&&
\nonumber{}
\hspace{-15mm}
+\frac{1/2\mu^{4-d}}{(d-4)(4\pi)^2}\left(\frac{\lambda_b}{4!}\right)^2
\eta_{\mu\nu}
\Bigl(
 [\phi^*\star\phi\star\phi^*\star\phi]
 +[\phi\star\phi\star\phi^*\star\phi^*]
\Bigr)
\\&&{}
\hspace{-15mm}
+\frac{\mu^{4-d}}{(d-4)(4\pi)^2}\frac{\lambda_a}{4!}\frac{\lambda_b}{4!}
\eta_{\mu\nu}
\Bigl(
 [\phi\star\phi^*\star\phi^*\star\phi]
 +[\phi^*\star\phi\star\phi\star\phi^*]
\Bigr).
\label{dfdf*=}
\end{eqnarray}

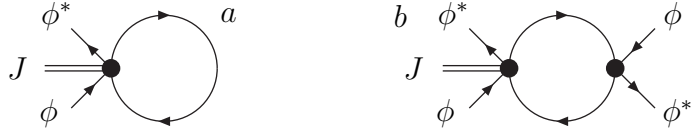
\begin{figure}[t]
\begin{picture}(450,50)(0,0)
\Vertex(130,20){3.5}
\ArrowArcn(150,20)(20,0,180)
\ArrowArcn(150,20)(20,180,360)
\Line(105,19)(130,19)
\Line(105,21)(130,21)
\ArrowLine(130,20)(115,35)
\ArrowLine(115,5)(130,20)
\Text(95,20)[c]{\mbox{}$J$}
\Text(110,40)[c]{\mbox{}$\phi^*$}
\Text(107,3)[c]{\mbox{}$\phi$}
\Text(175,40)[c]{$a$}
\Vertex(280,20){3.5}
\Vertex(320,20){3.5}
\ArrowArcn(300,20)(20,0,180)
\ArrowArcn(300,20)(20,180,360)
\Line(255,19)(280,19)
\Line(255,21)(280,21)
  \Text(245,20)[c]{\mbox{}$J$}
\ArrowLine(280,20)(265,35)
  \Text(260,40)[c]{\mbox{}$\phi^*$}
\ArrowLine(265,5)(280,20)
  \Text(257,3)[c]{\mbox{}$\phi$}
\ArrowLine(335,35)(320,20)
  \Text(343,40)[c]{\mbox{}$\phi$}
\ArrowLine(320,20)(335,5)
  \Text(345,3)[c]{\mbox{}$\phi^*$}
\Text(240,40)[c]{$b$}
\end{picture}
\caption{Divergent diagrams corresponding to the operator
$\phi^*\star\phi\star\phi^*\star\phi$}\label{ffff}
\end{figure}

Let us consider the renormalization of the operator
$\phi^*\star\phi\star\phi^*\star\phi$.
For this purpose we add to the action (\ref{action}) the
following term:
\begin{eqnarray}
\mu^{4-d}
\int d^dx\,
J\star\phi^*\star\phi\star\phi^*\star\phi
\label{Jf4}
\end{eqnarray}
and calculate all one-loop UV divergent contributions to the
effective action linear in $J$.
The diagrams which we need are shown in Figure~\ref{ffff}.
Also we notice that in eq.~(\ref{Jf4}) in order to keep the
source $J$ to be dimensionless, an arbitrary parameter $\mu$
with dimension of mass was introduced.
The UV divergences which arise from these diagrams are
\begin{eqnarray}
\frac{6m^2}{(d-4)(4\pi)^2}
\int d^4x\, J\star\phi^*\star\phi
\end{eqnarray}
for diagram $a$ and
\begin{eqnarray}
&&
\frac{\mu^{4-d}}{(d-4)(4\pi)^2}
\frac{\lambda_a}{2}
\int d^4x\, J\star\phi^*\star\phi\star\phi^*\star\phi
+
\\
&+&
\frac{\mu^{4-d}}{(d-4)(4\pi)^2}
\frac{2\lambda_b}{4!}
\int d^4x\,
J\star\Bigl[
\phi^*\star\phi^*\star\phi\star\phi
+
\phi^*\star\phi\star\phi\star\phi^*
+
\phi\star\phi^*\star\phi^*\star\phi
\Bigr]
\nonumber
\end{eqnarray}
for diagram $b$, respectively.
The result of the renormalization of the operator under
con\-si\-de\-ra\-tion is
\begin{eqnarray}
\nonumber
\frac{\lambda_{a0}}{4!}\,
\phi_0^*\star\phi_0\star\phi_0^*\star\phi_0
&=&
\mu^{4-d}
\left(\frac{\lambda_{a}}{4!}+\frac{1}{(d-4)(4\pi)^2}
       \frac{4\lambda_a^2-2\lambda_b^2}{(4!)^2}
\right)
 [\phi^*\star\phi\star\phi^*\star\phi]
\\&&\nonumber{}
\hspace{-3cm}
+\frac{2\mu^{4-d}}{(d-4)(4\pi)^2}\,\frac{\lambda_{a}}{4!}\frac{\lambda_b}{4!}
\Bigl(
[\phi^*\star\phi^*\star\phi\star\phi]
+[\phi^*\star\phi\star\phi\star\phi^*]
+[\phi\star\phi^*\star\phi^*\star\phi]
\Bigr)
\\&&{}
+\frac{1}{(d-4)(4\pi)^2}\,\frac{\lambda_{a}}{4!}\, 6m^2
[\phi^*\star\phi].
\label{f4a1}
\end{eqnarray}

Exchanging $\phi\leftrightarrow\phi^*$ in (\ref{f4a1}) we find the
renormalization of the operator
$\phi\star\phi^*\star\phi\star\phi^*$
\begin{eqnarray}
\nonumber
\frac{\lambda_{a0}}{4!}
\phi_0\star\phi_0^*\star\phi_0\star\phi_0^*
&=&
\mu^{4-d}
\left(\frac{\lambda_{a}}{4!}+\frac{1}{(d-4)(4\pi)^2}
       \frac{4\lambda_a^2-2\lambda_b^2}{(4!)^2}
\right)
 [\phi\star\phi^*\star\phi\star\phi^*]
\\&&\nonumber{}
\hspace{-3cm}
+\frac{2\mu^{4-d}}{(d-4)(4\pi)^2}\,\frac{\lambda_{a}}{4!}\,\frac{\lambda_b}{4!}
\Bigl(
[\phi\star\phi\star\phi^*\star\phi^*]
+[\phi\star\phi^*\star\phi^*\star\phi]
+[\phi^*\star\phi\star\phi\star\phi^*]
\Bigr)
\\&&
+\frac{1}{(d-4)(4\pi)^2}\,\frac{\lambda_{a}}{4!}\, 6m^2
[\phi\star\phi^*].
\end{eqnarray}

Carrying out analogous calculations, we may find the renormalization
relations for the remaining composite operators which enter into the
energy-momentum tensor (\ref{emt})
\begin{eqnarray}
\nonumber
\frac{\lambda_{b0}}{4!}
\phi_0^*\star\phi_0^*\star\phi_0\star\phi_0
&=&
\mu^{4-d}\frac{\lambda_{b}}{4!}
\left(1+\frac{1}{(d-4)(4\pi)^2}
   \frac{-4\lambda_a+2\lambda_b}{4!}
\right)
[\phi^*\star\phi^*\star\phi\star\phi]
\\&&\nonumber{}
+\frac{2\mu^{4-d}}{(d-4)(4\pi)^2}\left(\frac{\lambda_b}{4!}\right)^2
[\phi^*\star\phi\star\phi^*\star\phi]
\\&&{}
+\frac{1}{(d-4)(4\pi)^2}\,\frac{\lambda_{b}}{4!}\,2m^2 [\phi^*\star\phi],
\\[3mm]
\nonumber
\frac{\lambda_{b0}}{4!}\,
\phi_0\star\phi_0\star\phi_0^*\star\phi_0^*
&=&
\mu^{4-d}\frac{\lambda_{b}}{4!}
\left(1+\frac{1}{(d-4)(4\pi)^2}
   \frac{-4\lambda_a+2\lambda_b}{4!}
\right)
[\phi\star\phi\star\phi^*\star\phi^*]
\\&&\nonumber{}
+\frac{2\mu^{4-d}}{(d-4)(4\pi)^2}\left(\frac{\lambda_b}{4!}\right)^2
[\phi\star\phi^*\star\phi\star\phi^*]
\\&&{}
+\frac{1}{(d-4)(4\pi)^2}\,\frac{\lambda_{b}}{4!}\,2m^2 [\phi\star\phi^*],
\\[3mm]
\nonumber
\frac{\lambda_{b0}}{4!}\,
\phi_0^*\star\phi_0\star\phi_0\star\phi_0^*
&=&
\mu^{4-d}\,\frac{\lambda_{b}}{4!}
[\phi^*\star\phi\star\phi\star\phi^*]
\\&&
\nonumber{}
+\frac{2\mu^{4-d}}{(d-4)(4\pi)^2}\left(\frac{\lambda_b}{4!}\right)^2
\Bigl(
[\phi^*\star\phi\star\phi^*\star\phi]
+[\phi\star\phi^*\star\phi\star\phi^*]
\Bigr)
\\&&{}
+\frac{1}{(d-4)(4\pi)^2}\,\frac{\lambda_{b}}{4!}\,2m^2
\Bigl([\phi^*\star\phi]+[\phi\star\phi^*]\Bigr),
\\[3mm]
\nonumber
\frac{\lambda_{b0}}{4!}\,
\phi_0\star\phi_0^*\star\phi_0^*\star\phi_0
&=&
\mu^{4-d}\,\frac{\lambda_{b}}{4!}
[\phi\star\phi^*\star\phi^*\star\phi]
\\&&\nonumber{}
+\frac{2\mu^{4-d}}{(d-4)(4\pi)^2}\left(\frac{\lambda_b}{4!}\right)^2
\Bigl(
[\phi\star\phi^*\star\phi\star\phi^*]
+[\phi^*\star\phi\star\phi^*\star\phi]
\Bigr)
\\&&{}
+\frac{1}{(d-4)(4\pi)^2}\,\frac{\lambda_{b}}{4!}\,2m^2
\Bigl([\phi^*\star\phi]+[\phi\star\phi^*]\Bigr).
\label{last}
\end{eqnarray}
Using the renormalization relations which we have found in this
section we can define the mixing matrix $Z$ (\ref{Z}) in the
one-loop approximation.
In view of its huge size we write it
in explicit form in Table~\ref{Zmatrix} at page~\pageref{Zmatrix} of the
article.

Now we turn to the renormalization of the energy-momentum tensor
(\ref{emt}) of the theory.

\begin{landscape}
\begin{table}[p]
\begin{eqnarray}
Z&=&
\left(
\begin{array}{ccccccccccc}
1-\frac{1/2\lambda_a}{(4\pi)^2\varepsilon}&
  0&
  -\frac{2\lambda_b}{(4\pi)^24!\varepsilon}&
    0&
     -\frac{2\lambda_b}{(4\pi)^24!\varepsilon}&
       -\frac{6\mu^{-\varepsilon}}{(4\pi)^2\varepsilon}&
         0&0&0&0&0
\\
0&
 1-\frac{1/2\lambda_a}{(4\pi)^2\varepsilon}&
  0&
  -\frac{2\lambda_b}{(4\pi)^24!\varepsilon}&
     -\frac{2\lambda_b}{(4\pi)^24!\varepsilon}&
     0&
       -\frac{6\mu^{-\varepsilon}}{(4\pi)^2\varepsilon}&
         0&0&0&0
\\
-\frac{2\lambda_b}{(4\pi)^24!\varepsilon}&
  0&
   1-\frac{\lambda_a+\lambda_b}{(4\pi)^23!\varepsilon}&
    0&0&
     -\frac{2\mu^{-\varepsilon}}{(4\pi)^2\varepsilon}&
      0&0&0&0&0
\\
0&
 -\frac{2\lambda_b}{(4\pi)^24!\varepsilon}&
  0&
   1-\frac{\lambda_a+\lambda_b}{(4\pi)^23!\varepsilon}&
    0&0&
     -\frac{2\mu^{-\varepsilon}}{(4\pi)^2\varepsilon}&
      0&0&0&0
\\
-\frac{\lambda_b}{(4\pi)^23!\varepsilon}&
  -\frac{\lambda_b}{(4\pi)^23!\varepsilon}&
    0&0&
     1-\frac{8\lambda_a+2\lambda_b}{(4\pi)^24!\varepsilon}&
      -\frac{4\mu^{-\varepsilon}}{(4\pi)^2\varepsilon}&
        -\frac{4\mu^{-\varepsilon}}{(4\pi)^2\varepsilon}&
         0&0&0&0
\\
0&0&0&0&0&
1+\frac{\lambda_a+\lambda_b}{(4\pi)^23!\varepsilon}&
 -\frac{2\lambda_b}{(4\pi)^24!\varepsilon}&
   0&0&0&0
\\
0&0&0&0&0&
-\frac{2\lambda_b}{(4\pi)^24!\varepsilon}&
1+\frac{\lambda_a+\lambda_b}{(4\pi)^23!\varepsilon}&
   0&0&0&0
\\
0&0&0&0&0&0&0&
  1-\frac{\lambda_a}{(4\pi)^23!\varepsilon}&
   - \frac{2\lambda_b}{(4\pi)^24!\varepsilon}&
    0&0
\\
0&0&0&0&0&0&0&
 - \frac{2\lambda_b}{(4\pi)^24!\varepsilon}&
  1-\frac{\lambda_a}{(4\pi)^23!\varepsilon}&
    0&0
\\
0&0&0&0&0&0&0&0&0&1&0
\\
0&0&0&0&0&0&0&0&0&0&1
\end{array}
\right)
\nonumber
\\[5ex]
\nonumber
\\[5ex]
&&
\varepsilon=4-d
\nonumber
\end{eqnarray}
\caption{The mixing matrix in the one-loop
approximation}\label{Zmatrix}
\end{table}
\end{landscape}

\section{One-loop renormalization of the energy-mo\-men\-tum tensor}\label{EMT}

The purpose of this section is to construct the finite operator of the energy-momentum tensor of the theory under consideration.
Using the result of the previous section, we can see that the bare energy-momentum tensor
\begin{eqnarray}
T_{0\mu\nu}
&=&
c_1\biggl(
   \partial_\mu\phi_0^*\star\partial_\nu\phi_0
  +\partial_\nu\phi_0^*\star\partial_\mu\phi_0
 \biggr)
+
(1-c_1)
   \biggl(
   \partial_\mu\phi_0\star\partial_\nu\phi_0^*
  +\partial_\nu\phi_0\star\partial_\mu\phi_0^*
   \biggr)
\nonumber
\\&&{}
-\eta_{\mu\nu}\biggl(
   c_2\,\partial_\alpha\phi_0^*\star\partial^\alpha\phi_0
   +(1-c_2)\,\partial_\alpha\phi_0\star\partial^\alpha\phi_0^*
   \biggr)
\nonumber
\\&&{}
+\eta_{\mu\nu}\,m_0^2
   \biggl(
   c_3\,\phi_0^*\star\phi_0
   +(1-c_3)\,\phi_0\star\phi_0^*
      \biggr)
\nonumber
\\&&{}
+\eta_{\mu\nu}\,\frac{\lambda_{a0}}{4!}\,
\biggl(
c_4\, \phi_0^*\star\phi_0\star\phi_0^*\star\phi_0
+(1-c_4)\, \phi_0\star\phi_0^*\star\phi_0\star\phi_0^*
\biggr)
\nonumber
\\&&{}
+\eta_{\mu\nu}\,\frac{\lambda_{b0}}{4!}\,
\biggl(
c_5\, \phi_0^*\star\phi_0^*\star\phi_0\star\phi_0
+
c_6\, \phi_0\star\phi_0\star\phi_0^*\star\phi_0^*
\biggr)
\nonumber
\\&&{}
+\eta_{\mu\nu}\,\frac{\lambda_{b0}}{4!}\,\frac{1-c_5-c_6}{2} \biggl(\phi_0^*\star\phi_0\star\phi_0\star\phi_0^* + \phi_0\star\phi_0^*\star\phi_0^*\star\phi_0\biggr),
\label{bemt}
\end{eqnarray}
is not finite in the one-loop approximation.
Substituting the expressions which relate the bare operators
entering into the bare energy-momentum tensor with the
renormalized ones (\ref{first+}), (\ref{first}), (\ref{df*df=}),
(\ref{dfdf*=}) (\ref{f4a1}--\ref{last}) we
find its divergences
\begin{eqnarray}
\nonumber
T_{0\mu\nu}
&=&
[T_{\mu\nu}]
+
\\
\nonumber
&&{}
+
\partial^2_{\mu\nu}[\phi^*\star\phi]
\frac{1/3}{(d-4)(4\pi)^2\,4!}\,
\Bigl(
2(1-c_1)\lambda_b
+
4c_1\lambda_a
\Bigr)
\\
\nonumber
&&{}
+
\partial^2_{\mu\nu}[\phi\star\phi^*]
\frac{1/3}{(d-4)(4\pi)^2\,4!}
\Bigl(
2c_1\lambda_b
+
4(1-c_1)\lambda_a
\Bigr)
\\
\nonumber
&&{}
+
\eta_{\mu\nu}\partial^2[\phi^*\star\phi]
\frac{1/3}{(d-4)(4\pi)^2\,4!}
\Bigl(
\lambda_a(c_1-3c_2)
+
\lambda_b(-2-c_1+3c_2)
\Bigr)
\\
\nonumber
&&{}
+
\eta_{\mu\nu}\partial^2[\phi\star\phi^*]
\frac{1/3}{(d-4)(4\pi)^2\,4!}
\Bigl(
2\lambda_a(-2-c_1+3c_2)
+
\lambda_b(c_1-3c_2)
\Bigr)
\\&&{}
\nonumber
+
\eta_{\mu\nu}m^2[\phi^*\star\phi]
\frac{2}{(d-4)(4\pi)^2\,4!}
\Bigl(
\lambda_a(2c_1-4c_2-2c_3+3c_4)
\\
\nonumber
&&
\hspace{60mm}{}
+
\lambda_b(1-c_1+2c_2-3c_3-c_6)
\Bigr)
\\
&&{}
\nonumber
+
\eta_{\mu\nu}m^2[\phi\star\phi^*]
\frac{2}{(d-4)(4\pi)^2\,4!}
\Bigl(
\lambda_a(-1-2c_1+4c_2+2c_3-3c_4)
\\
\nonumber
&&{}
\hspace{60mm}{}
+
\lambda_b(-1+c_1-2c_2+3c_3-c_5)
\Bigr)
\\&&{}
\nonumber
+
\eta_{\mu\nu}
[\phi^*\star\phi\star\phi^*\star\phi]
\frac{\mu^{4-d}}{(d-4)(4\pi)^2(4!)^2}\times
\\
\nonumber
&&
\hspace{40mm}{}
\times
\Bigl(
4\lambda_a^2(c_1-2c_2+c_4)
+
\lambda_b^2(1-c_1+2c_2-2c_4-2c_6)
\Bigr)
\\&&{}
\nonumber
+
\eta_{\mu\nu}
[\phi\star\phi^*\star\phi\star\phi^*]
\frac{\mu^{4-d}}{(d-4)(4\pi)^2(4!)^2}\times
\\
\nonumber
&&{}
\hspace{40mm}
\times
\Bigl(
4\lambda_a^2(-c_1+2c_2-c_4)
+
\lambda_b^2(c_1-2c_2+2c_4-2c_5)
\Bigr)
\\
&&{}
\nonumber
+
\eta_{\mu\nu}
[\phi^*\star\phi^*\star\phi\star\phi]
\frac{\mu^{4-d}}{(d-4)(4\pi)^2}
\frac{\lambda_b}{(4!)^2}
\times
\\
&&{}
\nonumber
\hspace{50mm}
\times
\Bigl(
2\lambda_a(c_4-2c_5)
+
\lambda_b(c_1-2c_2+2c_5)
\Bigr)
\\&&{}
\nonumber
+
\eta_{\mu\nu}
[\phi\star\phi\star\phi^*\star\phi^*]
\frac{\mu^{4-d}}{(d-4)(4\pi)^2}
\frac{\lambda_b}{(4!)^2}
\times
\\
&&{}
\hspace{40mm}
\times
\Bigl(
\lambda_a(2-2c_4-4c_6)
+
\lambda_b(-1-c_1+2c_2+2c_6)
\Bigr).
\end{eqnarray}
Here $[T_{\mu\nu}]$ is a finite quantity
\begin{eqnarray}
[T_{\mu\nu}]
&=&
c_1\biggl(
   [\partial_\mu\phi^*\star\partial_\nu\phi]
  +[\partial_\nu\phi^*\star\partial_\mu\phi]
 \biggr)
+
(1-c_1)
   \biggl(
   [\partial_\mu\phi\star\partial_\nu\phi^*]
  +[\partial_\nu\phi\star\partial_\mu\phi^*]
   \biggr)
\nonumber
\\&&{}
-\eta_{\mu\nu}\biggl(
   c_2\,[\partial_\alpha\phi^*\star\partial^\alpha\phi]
   +(1-c_2)\,[\partial_\alpha\phi\star\partial^\alpha\phi^*]
   \biggr)
\nonumber
\\&&{}
+\eta_{\mu\nu}\,m^2
   \biggl(
   c_3\,[\phi^*\star\phi]
   +(1-c_3)\,[\phi\star\phi^*]
      \biggr)
\nonumber
\\&&{}
+\eta_{\mu\nu}\,\mu^{4-d}\,\frac{\lambda_a}{4!}\,
\biggl(
c_4\, [\phi^*\star\phi\star\phi^*\star\phi]
+(1-c_4)\, [\phi\star\phi^*\star\phi\star\phi^*]
\biggr)
\nonumber
\\&&{}
+\eta_{\mu\nu}\,\mu^{4-d}\,\frac{\lambda_b}{4!}\,
\biggl( c_5\, [\phi^*\star\phi^*\star\phi\star\phi]
+
c_6\, [\phi\star\phi\star\phi^*\star\phi^*]
\biggr)
\nonumber
\\&&{}
+\eta_{\mu\nu}\,\mu^{4-d}\,\frac{\lambda_b}{4!}\,\frac{1-c_5-c_6}{2} \biggl([\phi^*\star\phi\star\phi\star\phi^*] + [\phi\star\phi^*\star\phi^*\star\phi]\biggr).
\label{remt}
\end{eqnarray}

In order to make the energy-momentum tensor (\ref{bemt}) finite,
we add to it all possible real terms having the same mass dimensions and
symmetry with arbitrary coefficients, to be determined
\begin{eqnarray}
T_{0\mu\nu}
&+&
d_1\biggl(
   \partial_\mu\phi_0^*\star\partial_\nu\phi_0
  +\partial_\nu\phi_0^*\star\partial_\mu\phi_0
 \biggr)
+
d_2
   \biggl(
   \partial_\mu\phi_0\star\partial_\nu\phi_0^*
  +\partial_\nu\phi_0\star\partial_\mu\phi_0^*
   \biggr)
\nonumber
\\&&{}
+\eta_{\mu\nu}\,m_0^2
   \biggl(
   d_3\,\phi_0^*\star\phi_0
   +d_4\,\phi_0\star\phi_0^*
      \biggr)
\nonumber
\\&&{}
+
d_5\,\eta_{\mu\nu}\,\partial^2(\phi_0^*\star\phi_0)
+
d_6\,\eta_{\mu\nu}\,\partial^2(\phi_0\star\phi_0^*)
\nonumber
\\&&{}
+
d_7\biggl((\partial^2_{\mu\nu}\phi_0^*)\star\phi_0
+
\phi_0^*\star(\partial^2_{\mu\nu}\phi_0)
\biggr)
+
d_8\biggl((\partial^2_{\mu\nu}\phi_0)\star\phi_0^*
+
\phi_0\star(\partial^2_{\mu\nu}\phi_0^*)
\biggr)
\nonumber
\\&&{}
+
d_9\,\eta_{\mu\nu}
\biggl(
L^*_0\star\phi_0+\phi^*_0\star{}L_0
\biggr)
+
d_{10}\,\eta_{\mu\nu}
\biggl(
L_0\star\phi_0^*+\phi_0\star{}L_0^*
\biggr)
\nonumber
\\&&{}
+\eta_{\mu\nu}\,\frac{\lambda_{a0}}{4!}\,
\biggl(
d_{11}\, \phi_0^*\star\phi_0\star\phi_0^*\star\phi_0
+d_{12}\, \phi_0\star\phi_0^*\star\phi_0\star\phi_0^*
\biggr)
\nonumber
\\&&{}
+\eta_{\mu\nu}\,\frac{\lambda_{b0}}{4!}\,
\biggl(
d_{13}\, \phi_0^*\star\phi_0^*\star\phi_0\star\phi_0
+
d_{14}\, \phi_0\star\phi_0\star\phi_0^*\star\phi_0^*
\biggr)
\nonumber
\\&&{}
+\eta_{\mu\nu}\,\frac{\lambda_{b0}}{4!}\,d_{15} \biggl(\phi_0^*\star\phi_0\star\phi_0\star\phi_0^* + \phi_0\star\phi_0^*\star\phi_0^*\star\phi_0\biggr).
\label{addings}
\end{eqnarray}
Demanding that the expression (\ref{addings}) be finite in the one-loop approximation we find some restrictions on the arbitrary coefficients.
From these restrictions we can determine only some of them, while the others are still arbitrary.
Substituting the found coefficients back into (\ref{addings}) we get the general expression for the finite operator in
the energy-momentum tensor of noncommutative complex scalar field theory
\begin{eqnarray}
T^{\mbox{\it\tiny{}fin}}_{0\mu\nu}
&=&
c_1\biggl(
   \partial_\mu\phi_0^*\star\partial_\nu\phi_0
  +\partial_\nu\phi_0^*\star\partial_\mu\phi_0
 \biggr)
+
(1-c_1)
   \biggl(
   \partial_\mu\phi_0\star\partial_\nu\phi_0^*
  +\partial_\nu\phi_0\star\partial_\mu\phi_0^*
   \biggr)
\nonumber
\\&&{}
-\eta_{\mu\nu}\biggl(
   c_2\,\partial_\alpha\phi_0^*\star\partial^\alpha\phi_0
   +(1-c_2)\,\partial_\alpha\phi_0\star\partial^\alpha\phi_0^*
   \biggr)
\nonumber
\\&&{}
+\eta_{\mu\nu}\,m_0^2
   \biggl(
   (c_2-c_1/2)\,\phi_0^*\star\phi_0
   +(1/2+c_1/2-c_2)\,\phi_0\star\phi_0^*
      \biggr)
\nonumber
\\&&{}
+\eta_{\mu\nu}\,\frac{\lambda_{a0}}{4!}\,
\biggl(
(2c_2-c_1)\, \phi_0^*\star\phi_0\star\phi_0^*\star\phi_0
+(1+c_1-2c_2)\, \phi_0\star\phi_0^*\star\phi_0\star\phi_0^*
\biggr)
\nonumber
\\&&{}
+\eta_{\mu\nu}\,\frac{\lambda_{b0}}{4!}\,
\biggl(
(c_2-c_1/2)\, \phi_0^*\star\phi_0^*\star\phi_0\star\phi_0
+
(1/2+c_1/2-c_2)\, \phi_0\star\phi_0\star\phi_0^*\star\phi_0^*
\biggr)
\nonumber
\\&&{}
+\eta_{\mu\nu}\,\frac{\lambda_{b0}}{4!}\,
\frac{1}{4}
\biggl(
  \phi_0^*\star\phi_0\star\phi_0\star\phi_0^*
  +
  \phi_0\star\phi_0^*\star\phi_0^*\star\phi_0
\biggr)
\nonumber
\\&&{}
+
\frac{2c_2-c_1}{3}\,
(\eta_{\mu\nu}\partial^2-\partial^2_{\mu\nu})(\phi_0^*\star\phi_0)
+
\frac{1+c_1-2c_2}{3}\,
(\eta_{\mu\nu}\partial^2-\partial^2_{\mu\nu})(\phi_0\star\phi_0^*)
\nonumber
\\&&{}
-
2(c_1-c_2)/3\,
\biggl(
\partial^2_{\mu\nu}-1/4\eta_{\mu\nu}\partial^2
\biggr)\,
(\phi_0^*\star\phi_0-\phi_0\star\phi_0^*)
\nonumber
\\&&{}
+
\Bigl(
d_9-(3d_1+c_1)/8
\Bigr)\,
\eta_{\mu\nu}\,
\biggl(
\phi_0^*\star{}L_0+L_0^*\star\phi_0
\biggr)
\nonumber
\\&&{}
+
\Bigl(
d_{10}-(3d_2+1-c_1)/8
\Bigr)\,
\eta_{\mu\nu}\,
\biggl(
\phi_0\star{}L_0^*+L_0\star\phi_0^*
\biggr)
\nonumber
\\&&{}
+
\Bigl(
3d_1+c_1
\Bigr)\,
S_{10\mu\nu}
+
\Bigl(
3d_2+1-c_1
\Bigr)\,
S_{20\mu\nu},
\label{igemt}
\end{eqnarray}
where we have introduced the following notation for the
traceless operators which are finite in the one-loop
approximation:
\begin{eqnarray}
S_{10\mu\nu}
&=&
 \frac{1}{2}
   \biggl(
     \partial_\mu\phi_0^*\star\partial_\nu\phi_0
     +
     \partial_\nu\phi_0^*\star\partial_\mu\phi_0
   \biggr)
-\frac{1}{6}\partial^2_{\mu\nu}(\phi_0^*\star\phi_0)
\nonumber
\\&&\qquad{}
-\frac{1}{4}\eta_{\mu\nu}\,\partial_\alpha\phi_0^*\star\partial^\alpha\phi_0
+\frac{1}{24}\eta_{\mu\nu}\,\partial^2(\phi_0^*\star\phi_0),
\label{S1}
\\
S_{20\mu\nu}
&=&
 \frac{1}{2}
   \biggl(
     \partial_\mu\phi_0\star\partial_\nu\phi_0^*
     +
     \partial_\nu\phi_0\star\partial_\mu\phi_0^*
   \biggr)
-\frac{1}{6}\partial^2_{\mu\nu}(\phi_0\star\phi_0^*)
\nonumber
\\&&\qquad{}
-\frac{1}{4}\eta_{\mu\nu}\,\partial_\alpha\phi_0\star\partial^\alpha\phi_0^*
+\frac{1}{24}\eta_{\mu\nu}\,\partial^2(\phi_0\star\phi_0^*).
\label{S2}
\end{eqnarray}
Let us notice that the ordering coefficients $c_3$, $c_4$, $c_5$, $c_6$ do not enter
into the general expression for the finite energy-momentum tensor (\ref{igemt}) of the theory.
As far as the undefined arbitrary coefficients $d_1$, $d_2$, $d_9$, $d_{10}$ are concerned, the operators
standing after them are finite in the one-loop approximation.
For simplicity we make the coefficients standing before these finite operators to be zero by choosing
$d_1$, $d_2$, $d_9$, $d_{10}$ in a proper way.
The resulting "improved" energy-momentum tensor is
\begin{eqnarray}
T^{I}_{0\mu\nu}
&=&
c_1\biggl(
   \partial_\mu\phi_0^*\star\partial_\nu\phi_0
  +\partial_\nu\phi_0^*\star\partial_\mu\phi_0
 \biggr)
+
(1-c_1)
   \biggl(
   \partial_\mu\phi_0\star\partial_\nu\phi_0^*
  +\partial_\nu\phi_0\star\partial_\mu\phi_0^*
   \biggr)
\nonumber
\\&&{}
-\eta_{\mu\nu}\biggl(
   c_2\,\partial_\alpha\phi_0^*\star\partial^\alpha\phi_0
   +(1-c_2)\,\partial_\alpha\phi_0\star\partial^\alpha\phi_0^*
   \biggr)
\nonumber
\\&&{}
+\eta_{\mu\nu}\,m_0^2
   \biggl(
   (c_2-c_1/2)\,\phi_0^*\star\phi_0
   +(1/2+c_1/2-c_2)\,\phi_0\star\phi_0^*
      \biggr)
\nonumber
\\&&{}
+\eta_{\mu\nu}\,\frac{\lambda_{a0}}{4!}\,
\biggl(
(2c_2-c_1)\, \phi_0^*\star\phi_0\star\phi_0^*\star\phi_0
+(1+c_1-2c_2)\, \phi_0\star\phi_0^*\star\phi_0\star\phi_0^*
\biggr)
\nonumber
\\&&{}
+\eta_{\mu\nu}\,\frac{\lambda_{b0}}{4!}\,
\biggl(
(c_2-c_1/2)\, \phi_0^*\star\phi_0^*\star\phi_0\star\phi_0
+
(1/2+c_1/2-c_2)\, \phi_0\star\phi_0\star\phi_0^*\star\phi_0^*
\biggr)
\nonumber
\\&&{}
+\eta_{\mu\nu}\,\frac{\lambda_{b0}}{4!}\,
\frac{1}{4}
\biggl(
  \phi_0^*\star\phi_0\star\phi_0\star\phi_0^*
  +
  \phi_0\star\phi_0^*\star\phi_0^*\star\phi_0
\biggr)
\nonumber
\\
&&{}
+
\frac{2c_2-c_1}{3}\,
(\eta_{\mu\nu}\partial^2-\partial^2_{\mu\nu})(\phi_0^*\star\phi_0)
+
\frac{1+c_1-2c_2}{3}\,
(\eta_{\mu\nu}\partial^2-\partial^2_{\mu\nu})(\phi_0\star\phi_0^*)
\nonumber
\\
&&{}
-
2(c_1-c_2)/3\,
\biggl(
\partial^2_{\mu\nu}-1/4\eta_{\mu\nu}\partial^2
\biggr)\,
(\phi_0^*\star\phi_0-\phi_0\star\phi_0^*).
\label{iemt}
\end{eqnarray}
If we consider the commutative limit $\theta^{\mu\nu}\to{}0$ of
this expression, we get the improved energy-momentum
tensor of the commutative complex scalar field theory up to the
mass term.
Let us also notice that the expression for the "improved"
energy-momentum tensor (\ref{iemt}) is traceless unlike the
commutative case.
This situation is completely analogous to the case of real field
theory \cite{0303186}: in the noncommutative case the "improved"
energy-momentum is traceless, and in the commutative limit it
coincides with the "improved" energy-momentum tensor of the
corresponding commutative theory up to the mass term.

Let us check if the "improved" energy-momentum tensor
(\ref{iemt}) leads to global conserved quantities.
For this end we calculate its divergence
\begin{eqnarray}
\partial^\nu T_{0\mu\nu}^I
&=&
c_1\,\partial^\nu
\biggl\{
  \partial_\mu\phi_0^*,\partial_\nu\phi_0
\biggr\}
+
c_1\,\partial^\nu
\biggl\{
  \partial_\nu\phi_0^*,\partial_\mu\phi_0
\biggr\}
+
c_2\,\partial_\mu
\biggl\{
  \partial_\alpha\phi_0^*,\partial^\alpha\phi_0
\biggr\}
\nonumber
\\&&{}
+
m_0^2\,\Bigl(c_2-c_1/2\Bigr)\,
\partial_\mu
\biggl\{
\phi_0^*,\phi_0
\biggr\}
+
\frac{1}{4}
\Bigl(c_2-c_1\Bigr)\,
\partial_\mu\partial^2
\biggl\{
\phi_0^*,\phi_0
\biggr\}
\nonumber
\\&&{}
+
\frac{\lambda_{a0}}{4!}
\Bigl(2c_2-c_1\Bigr)\,
\partial_\mu
\biggl\{
\phi_0^*,\phi_0\star\phi_0^*\star\phi_0
\biggr\}
+
\frac{\lambda_{b0}}{4!}
\Bigl(c_2-c_1/2\Bigr)\,
\partial_\mu
\biggl\{
\phi_0^*\star\phi_0^*,\phi_0\star\phi_0
\biggr\}
\nonumber
\\&&{}
+
\frac{\lambda_{a0}}{4!}
\Bigg(
\biggl\{
\phi_0\star\phi_0^*,(\partial_\mu\phi_0)\star\phi_0^*
\biggr\}
+
\biggl\{
\phi_0\star\partial_\mu\phi_0^*,\phi_0\star\phi_0^*
\biggr\}
\Biggr)
\nonumber
\\&&{}
+
\frac{\lambda_{b0}}{4!}
\Biggl(
\biggl\{
\phi_0\star\phi_0\star\partial_\mu\phi_0^*,\phi_0^*
\biggr\}
+
\biggl\{
\phi_0,(\partial_\mu\phi_0)\star\phi_0^*\star\phi_0^*
\biggr\}
\Biggr)
\nonumber
\\&&{}
+
\frac{1}{4}
\frac{\lambda_{b0}}{4!}\,
\partial_\mu
\Biggl(
\biggl\{
\phi_0\star\phi_0^*\star\phi_0^*,\phi_0
\biggr\}
+
\biggl\{
\phi_0^*,\phi_0\star\phi_0\star\phi_0^*
\biggr\}
\Biggr)
\nonumber
\\&&{}
-\frac{m_0^2}{2}\partial_\mu(\phi_0\star\phi_0^*),
\label{div}
\end{eqnarray}
with $\{A,B\}=A\star{}B-B\star{}A$ being the Moyal bracket.
In the case of spatial noncommutativity $\theta^{0i}=0$
the Moyal bracket is a spatial divergence
$\{A,B\}=\partial^iC_i$, with $C_i$ being some functions of the
fields of the theory.
So, the expression (\ref{div}) is a spatial divergence only in the
case when $\mu$ is a spatial index because of the last line.
In this case after integration over space coordinates we have
$\int{}\partial^\nu{}J_{i\nu}\,d^3x=0$, and $J_{i0}$ are
conserved.
The quantity $J_{00}$ is conserved if the mass of the field is
zero.
The same situation occurs in the noncommutative theory of a real
scalar field \cite{0303186}: the energy of the field is
conserved in the massless case only.

It is interesting to note that the "improved" energy-momentum
tensor (\ref{iemt}) depends on the two arbitrary ordering
coefficients $c_1$ and $c_2$ which do not influence its
renormalizability, tracelessness and conservation conditions.

\section{Renormalization of the composite operators at zero
momentum transfer}\label{NCComOp0}

The purpose of this section is to renormalize scalar
hermitian composite operators of the theory under
consideration at zero momentum transfer.
As we have seen in the previous section (see e.g. (\ref{f*fa}),
(\ref{f*fb})), in noncommutative field theories the UV
divergence of a diagram depends on the value of external
momenta, so we need different counterterms if the momentum transfer
($k$ in our formula)  is equal to zero.
This situation is typical for any noncommutative field theory
(see the discussion of this problem in noncommutative real scalar
field theory in \cite{0303186}).
Since we may expand any operator on some basis it is sufficient
to study renormalization of those operators which constitute a
basis.
We may take as a basis the operators (\ref{bases}) which are
integrated over the whole space-time.
However, because of the cyclic property (\ref{cycle}) the number of
independent operators is reduced.
Also the operators which are a total divergence
disappear when we integrate them over the whole space-time.
We choose the following operators as a basis:
\begin{eqnarray}
Q^{(0)}_0=\left(
\begin{array}{c}
\int d^dx\,
\phi_0^*\star\phi_0\star\phi_0^*\star\phi_0\\
\int d^dx\,
\phi_0^*\star\phi_0^*\star\phi_0\star\phi_0\\
m_0^2\int d^dx\,\phi_0^*\star\phi_0\\
\int d^dx\,
(\phi_0^*\star{}L_0+L_0^*\star\phi_0)
\end{array}
\right)
&\mbox{and}&
[Q^{(0)}]=\left(
\begin{array}{c}
[\int d^dx\,\phi^*\star\phi\star\phi^*\star\phi]\\{}
[\int d^dx\,\phi^*\star\phi^*\star\phi\star\phi]\\{}
m^2\,[\int d^dx\,\phi^*\star\phi]\\{}
[\int d^dx\,(\phi^*\star{}L+L^*\star\phi)]
\end{array}
\right)
\label{bases'}
\end{eqnarray}
for the bare and renormalized operators respectively.
We see that the number of independent operators is
reduced in comparison with the case of arbitrary momentum
transfer (\ref{bases}).
Composite operators at zero momentum transfer are also mixed
by renormalization and we may write
\begin{eqnarray}
Q_0^{(0)}&=&Z^{(0)}[Q^{(0)}],
\label{Z0}
\end{eqnarray}
with $Z^{(0)}$ being a mixing matrix for the basis of composite operators
at zero momentum transfer (\ref{bases'}) which usually differs
from $Z$ in (\ref{Z}) in noncommutative field theories.

In this section we calculate this mixing matrix $Z^{(0)}$ in the
one-loop approximation.

Let us consider the renormalization of the operator
$m^2\,\int{}d^dx\,\phi_0^*\star\phi_0$.
This operator may be obtained from (\ref{Jff}) taking $J(x)=1$ or
$\tilde{J}(k)=(2\pi)^d\delta(k)$ in momentum space.
By dimensional analysis we may show that the same diagrams as in
the case of arbitrary momentum transfer may contain UV
divergences.
They are shown in Figure~\ref{f*f}.
But unlike that case, now we have a vanishing external momentum $k$.
Therefore (see the discussion after formulae (\ref{f*fb})), in
addition to integrals (\ref{f*fa}), (\ref{f*fc}), also integrals
(\ref{f*fb}), (\ref{f*ff}) become UV divergent.
As a result the UV divergences of the diagrams shown in
Figure~\ref{f*f} are
\begin{eqnarray}
&&\frac{1}{(d-4)(4\pi)^2}\,\frac{\lambda_a}{3}\,\,m^2
\int d^4x\,\phi^*\star\phi,
\\
&&\frac{1}{(d-4)(4\pi)^2}\,\frac{\lambda_b}{3!}\,\,m^2
\int d^4x\,\phi\star\phi^*,
\end{eqnarray}
and we have the following renormalization relation for the
operator under consideration
\begin{eqnarray}
m_0^2\int d^dx\, \phi_0^*\star\phi_0
&=&
m^2[\int d^4x\,\phi^*\star\phi].
\end{eqnarray}
This renormalization relation is similar to that of the
commutative field theory (\ref{comff}).
The same situation occurs in the theory of real scalar field
theory:
the renormalization relation for composite operators at zero
momentum transfer in the noncommutative theory is similar to
the corresponding renormalization relation in the commutative
theory \cite{0303186}.

As far as the remaining composite operators in (\ref{bases'}) are
concerned, the same situation occurs.
Since one of the external momenta is zero
the number of UV divergent diagrams becomes bigger.
The renormalization relation for the composite
operators at zero momentum transfer changes, in comparison with the case
of arbitrary momentum transfer
\begin{eqnarray}
\int d^dx\, \phi_0^*\star\phi_0\star\phi_0^*\star\phi_0
&=&
\left(1+\frac{2/3\lambda_a}{(d-4)(4\pi)^2}\right)\,
[\int d^dx\, \phi^*\star\phi\star\phi^*\star\phi]
\nonumber
\\
&&{}
+\frac{1/3\lambda_b}{(d-4)(4\pi)^2}\,
[\int d^dx\, \phi^*\star\phi^*\star\phi\star\phi]
\nonumber
\\
&&{}
+\frac{8\mu^{d-4}}{(d-4)(4\pi)^2}\,
m^2[\int d^dx\, \phi^*\star\phi^*],
\\
\int d^dx\, \phi_0^*\star\phi_0^*\star\phi_0\star\phi_0
&=&
\left(1+\frac{2\lambda_a+\lambda_b}{(d-4)(4\pi)^23!}\right)\,
[\int d^dx\, \phi^*\star\phi^*\star\phi\star\phi]
\nonumber
\\
&&{}
+\frac{1/6\lambda_b}{(d-4)(4\pi)^2}\,
[\int d^dx\, \phi^*\star\phi\star\phi^*\star\phi]
\nonumber
\\
&&{}
+\frac{4\mu^{d-4}}{(d-4)(4\pi)^2}\,
m^2[\int d^dx\, \phi^*\star\phi^*].
\end{eqnarray}
As a result we have that the mixing matrix $Z^{(0)}$ in the one-loop
approximation is
\begin{eqnarray}
Z^{(0)}&=&
\left(
\begin{array}{cccc}
1+\frac{2/3\lambda_a}{(d-4)(4\pi)^2}&
   \frac{1/3\lambda_b}{(d-4)(4\pi)^2}&
     \frac{8\mu^{d-4}}{(d-4)(4\pi)^2}&0\\
\frac{1/6\lambda_b}{(d-4)(4\pi)^2}&
  1+\frac{2\lambda_a+\lambda_b}{(d-4)(4\pi)^23!}&
     \frac{8\mu^{d-4}}{(d-4)(4\pi)^2}&0\\
0&0&1&0\\
0&0&0&1\\
\end{array}
\right).
\end{eqnarray}

In the next section we study the energy-momentum vector which
follows from the Noether's procedure.

\section{Renormalization of the energy-momentum vector}\label{EMV}
The purpose of this section is to construct the
finite and conserved energy-momentum vector of noncommutative complex scalar field theory.
From expression (\ref{Noether}) we may define it as
\begin{eqnarray}
\nonumber
P_\mu
&=&
\int
\biggl(
   \partial_\mu\phi_0^*\star\partial_0\phi_0
  +\partial_0\phi_0^*\star\partial_\mu\phi_0
  -\eta_{0\mu}
          \partial_\alpha\phi_0^*\star{}\partial^\alpha\phi_0
+\eta_{0\mu}
    \,m_0^2\phi_0^*\star{}\phi_0
\\&&\qquad{}
+\eta_{0\mu}
   \frac{\lambda_{a0}}{4!}\,\phi_0^*\star\phi_0\star\phi_0^*\star\phi_0
+\eta_{0\mu}
   \frac{\lambda_{b0}}{4!}\,\phi_0^*\star\phi_0^*\star\phi_0\star\phi_0
\biggr)
\, d^{d-1}x\, .
\label{emv}
\end{eqnarray}
As we noted above,
since $\theta^{0i}=0$, we have no time derivatives in the star product (\ref{theta}), and consequently, the properties
(\ref{cycle2}) and (\ref{cycle}) are still valid in the case of spatial integration only.
Therefore we have no problem of field ordering, and the energy-momentum vector is defined unambiguously.
It is evident that $P_\mu$ is conserved in time $\partial^0P_\mu=0$.
Now we show that this operator is finite, at least in the one-loop approximation.

In order to prove the finiteness of the energy-momentum vector (\ref{emv}), we
need to renormalize each of the six composite operators appearing in its expression.
As an example, let us consider the renormalization of the operator $m_0^2\int{}\phi_0^*\star\phi_0\,d^{d-1}x$.
In order to renormalize such an operator, we put $J(x)=\delta(x^0-t)$ in the integrand (\ref{Jff}).
In momentum space we have that $J\sim\delta(\vec{k})$.
In the case of spatial noncommutativity this leads to $\theta^{\mu\nu}k_\nu=0$
and, apart from (\ref{f*fa}) and (\ref{f*fc}), also expressions
(\ref{f*fb}) and (\ref{f*ff}) become UV divergent.
This is the appearence of the UV/IR mixing:
the divergences of a diagram depends on whether we put some of external momenta to zero before or after regularization is removed.
As a result we have the following renormalization relation
\begin{eqnarray}
m_0^2\int\phi_0^*\star\phi_0\,d^{d-1}x
=
m^2\biggl[\int\phi^*\star\phi\,d^3x\biggr].
\label{first2}
\end{eqnarray}
For the remaining composite operators similar calculations give
\begin{eqnarray}
\int\partial_\mu\phi_0^*\star\partial_\nu\phi_0\,d^{d-1}x
&=&
\Biggl[\int\partial_\mu\phi^*\star\partial_\nu\phi\,d^3x\Biggr]
\nonumber{}
\\&&
\hspace{-35mm}
\nonumber{}
+\frac{2}{(d-4)(4\pi)^2}\frac{2\lambda_a+\lambda_b}{4!}
\left(\frac{1}{6}\eta_{\mu\nu}\partial^2_{00}
     +\frac{1}{3}\delta_\mu^0\delta_\nu^0\partial^2_{00}
     +\eta_{\mu\nu}\,m^2
\right)
\Biggl[\int\phi^*\star\phi\,d^3x\Biggr]
\\&&
\hspace{-20mm}
\nonumber{}
+\frac{\mu^{4-d}}{(d-4)(4\pi)^2}
\frac{4\lambda_a^2+\lambda_b^2}{(4!)^2}\,
\eta_{\mu\nu}\Biggl[\int\phi^*\star\phi\star\phi^*\star\phi\,d^3x\Biggr]
\\&&{}
\hspace{-20mm}
+\frac{\mu^{4-d}}{(d-4)(4\pi)^2}
\frac{\lambda_b}{4!}\frac{4\lambda_a+\lambda_b}{4!}\,
\eta_{\mu\nu}
\Biggl[\int\phi^*\star\phi^*\star\phi\star\phi\,d^3x\Biggr],
\\[3mm]
\nonumber
\frac{\lambda_{a0}}{4!}\,
\int\phi_0^*\star\phi_0\star\phi_0^*\star\phi_0\,d^{d-1}x
&=&
\\&&
\nonumber{}
\hspace{-20mm}
=\mu^{4-d}
\left(\frac{\lambda_{a}}{4!}+\frac{1}{(d-4)(4\pi)^2}
   \frac{8\lambda_a^2-2\lambda_b^2}{(4!)^2}
\right)
 \Biggl[\int\phi^*\star\phi\star\phi^*\star\phi\,d^3x\Biggr]
\\&&
\nonumber{}
\hspace{-20mm}
+\frac{8\mu^{4-d}}{(d-4)(4\pi)^2}\,\frac{\lambda_{a}}{4!}\frac{\lambda_b}{4!}
\Biggl[\int\phi^*\star\phi^*\star\phi\star\phi\,d^3x\Biggr]
\\&&{}
\hspace{-20mm}
+\frac{8}{(d-4)(4\pi)^2}\,\frac{\lambda_{a}}{4!}\, m^2 \Biggl[\int\phi^*\star\phi\,d^3x\Biggr],
\\[3mm]
\nonumber
\frac{\lambda_{b0}}{4!}
\int\phi_0^*\star\phi_0^*\star\phi_0\star\phi_0\,d^{d-1}x
&=&
\mu^{4-d}\frac{\lambda_{b}}{4!}
\left(1+\frac{2}{(d-4)(4\pi)^2}
   \frac{\lambda_{b}}{4!}
\right)
\Biggl[\int\phi^*\star\phi^*\star\phi\star\phi\,d^3x\Biggr]
\\&&
\hspace{-20mm}
\nonumber{}
+\frac{4\mu^{4-d}}{(d-4)(4\pi)^2}\left(\frac{\lambda_b}{4!}\right)^2
\Biggl[\int\phi^*\star\phi\star\phi^*\star\phi\,d^3x\Biggl]
\\&&{}
\hspace{-20mm}
+\frac{4}{(d-4)(4\pi)^2}\,\frac{\lambda_{b}}{4!}\,m^2 \Biggl[\int\phi^*\star\phi\,d^3x\Biggr].
\label{last2}
\end{eqnarray}
Substituting (\ref{first2}--\ref{last2}) in (\ref{emv}) we find
that the energy-momentum vector (\ref{emv}) is finite in the
one-loop approximation
\begin{eqnarray}
\nonumber
P_\mu
&=&
\Biggl[\int
   \partial_\mu\phi^*\star\partial_0\phi\,d^3x
\Biggr]
+
\Biggl[\int
\partial_0\phi^*\star\partial_\mu\phi\,d^3x
\Biggr]
-
\eta_{0\mu}
\Biggl[\int
   \partial_\alpha\phi^*\star{}\partial^\alpha\phi\,d^3x
\Biggr]
\\&&{}
\nonumber
+
\eta_{0\mu}\,m^2
\Biggl[\int
\phi^*\star{}\phi\,d^3x
\Biggr]
+
\eta_{0\mu}
   \frac{\mu^{4-d}\lambda_a}{4!}\,
\Biggl[\int
   \phi^*\star\phi\star\phi^*\star\phi\,d^3x
\Biggr]
\\&&{}
+
\eta_{0\mu}
   \frac{\mu^{4-d}\lambda_b}{4!}\,
\Biggl[\int
   \phi^*\star\phi^*\star\phi\star\phi\,d^3x
\Biggr].
\label{remv}
\end{eqnarray}
This situation is similar to that in the noncommutative scalar field theory \cite{0303186}:
the energy-momentum vectors which follow from the Noether's theorem are finite in both theories and do not require improving,
but the energy-momentum tensors must be improved in order to be
finite, and this improving makes them conserved only in the massless
case.

\section{Summary}\label{Summary}
In this paper we have derived with the help of the Noether's
procedure the classical energy-momentum tensor of the
noncommutative complex scalar field theory.
It was shown that it cannot be defined unambigously and its
expression is defined up to six arbitrary ordering constants.

Next we have considered the renormalization of dimension four
composite operators of the theory and have found that the
renormalization of any composite operators of the theory demands
to take into account all composite operators with the same mass
dimension.
This phenomenon is called operator mixing and is typical for the
renormalization of composite operators of any theory.
The proper bases of hermitian scalar operators have been
constructed both for the bare and renormalized operators.
Due to the noncommutativity the number of the operators in the
bases is larger, in comparison with the commutative theory.
The mixing matrix which expresses the bare operators of the
basis in term of the renormalized ones is calculated in the one-loop
approximation.

We considered the renormalization of the energy-momentum tensor which
follows from the Noether's theorem and found it to be divergent
in the one-loop approximation.
In order to make it finite we have to add "improving" terms to
it.
The expression for the "improved" energy-momentum tensor has
been calculated and shown to be, apart from
traceless, conserved in the massless case only.
Besides, some ordering constants do not enter into the
expression for the "improved" energy-momentum tensor.

The renormalization of the composite operators at zero momentum
transfer was also considered.
The number of operators in the bases of such operators is
reduced in comparison with the case of arbitrary momentum
transfer, although it is bigger than in the corresponding commutative case
due to the noncommutativity.
The mixing matrix for the case of zero momentum transfer was
calculated in the one-loop approximation.
Finally we find that, as in the case of noncommutative real scalar
field theory \cite{0303186}, the energy-momentum vector which
follows from the Noether's theorem is conserved and finite in
the one-loop approximation.

\section*{Acknowledgements}
This work was supported in part by
the INTAS grant, project No.\ 03-51-6346,
the European Community's Human Potential
Programme under
contract HPRN-CT-2000-00131 Quantum Spacetime,
the INTAS-00-00254 grant
and the NATO Collaborative Linkage Grant PST.CLG.979389.
The work of I.L.B. and V.A.K. was also supported by
the RFBR grant, project No.\ 03-02-16193,
the joint RFBR-DFG grant, project No.\ 02-02-04002,
the DFG grant, project No.\ 436 RUS 113/669,
the grant for LSS, project No.\ 1252.2003.2
and
the grant PD02-1{.}2-94 of Russian Ministry of Education.
I.L.B. and V.A.K. wish to thank the Humboldt-Universit\"at zu
Berlin, where part of this work was done
and  D. L\"ust for warm hospitality.
I.L.B. and V.A.K
are also grateful for partial support to INFN, Laboratori
Nazionali di Frascati where the work was initiated.


\end{document}